\documentclass[aps,prd,reprint,amsmath,amssymb,longbibliography,superscriptaddress,floatfix]{revtex4-2}
\usepackage[T1]{fontenc}
\usepackage{graphicx}
\usepackage{bm}
\usepackage{booktabs}
\usepackage{xcolor}
\usepackage{hyperref}
\hypersetup{colorlinks=true,linkcolor=blue!55!black,citecolor=blue!55!black,urlcolor=blue!55!black,
pdfauthor={Dustin Keller},
pdftitle={Symmetry-adapted tensor networks and quantum-classical geometric learning for hadronic tomography: a proton construction}}
\newcommand{\F}{F_{1,1}^{u-\bar u}}
\newcommand{\Ft}{\widetilde F_{1,1}^{u-\bar u}}
\newcommand{\MS}{\overline{\mathrm{MS}}}
\newcommand{\GeV}{\mathrm{GeV}}
\newcommand{\Tr}{\operatorname{Tr}}
\newcommand{\dd}{\mathrm{d}}
\newcommand{\ii}{\mathrm{i}}
\begin{document}
\title{Symmetry-adapted tensor networks and quantum--classical geometric learning for hadronic tomography: a proton construction}
\author{Dustin Keller}
\affiliation{University of Virginia, Charlottesville, Virginia, USA}
\date{\today}
\begin{abstract}
We construct a calibrated model of the proton generalized transverse-momentum-dependent distribution (GTMD) $F_{1,1}$ in the real, staple-even valence-$u$ sector at zero skewness. Experimental measurements, lattice matrix elements, and phenomenological distributions enter through a single shared light-front state, organized as a symmetry-resolved tree tensor network in which color-singlet structure, exchange statistics, and the additive $J_z$ and $I_z$ charges are built in. Classical and quantum geometric nodes represent the residual transverse response. The GTMD is defined with soft subtraction, one-loop nonsinglet matching, NLO collinear evolution, and an explicit nonperturbative model for the joint dependence on quark transverse momentum and momentum transfer (the joint closure), so that each input is used with its own operator convention and scale. Calibrating against correlated transverse-momentum projections of the ART25 TMD extraction, HERA inclusive data, and lattice, elastic, and static information gives a simple picture. The forward inputs narrow the attainable transverse response by about $81\%$ at $x=0.4$. The simultaneous dependence on quark momentum and momentum transfer, however, is untouched by any of them: a deformation that leaves every present input invariant moves the nonforward value from $0.588$ to $0.378\,\GeV^{-2}$. Dense momentum-space surfaces and finite-transfer-window phase-space maps show the structure this freedom controls. The tensor-network organization is chosen with composition in mind. The same shared subunits, operator hierarchy, and validation protocol are designed to extend to the neutron, the spin-1 deuteron, and spin-$3/2$ nuclei, where the number of symmetry and exchange constraints grows quickly and a structural representation becomes the practical route. The quantum circuits serve as feature maps and state encodings and are checked against exact state-vector contractions; their capacity and cost are measured alongside the classical blocks.
\end{abstract}
\maketitle

\section{Introduction}
\label{sec:introduction}
Generalized transverse-momentum-dependent distributions (GTMDs) contain, as projections, the transverse-momentum-dependent distributions, generalized parton distributions, form factors, and collinear distributions of a hadron~\cite{meissner}, and their Fourier transforms in the transverse momentum transfer give the Wigner phase-space picture~\cite{lorce}. The information that bears on them is correspondingly scattered. Inclusive scattering constrains one projection, elastic observables another, lattice matrix elements a third, and a phenomenological TMD extraction a fourth. One would like to combine all of this into a single hadronic model in which each input constrains what it actually measures, the result can be read out in any variable, and whatever the inputs do not determine is isolated and quantified instead of being filled in by an ansatz.

We do this with a shared-state construction. A finite light-front proton state and a set of effective currents are organized as a symmetry-resolved tree tensor network (TTN), and every observable we use, from a DIS structure function to a lattice moment to the GTMD itself, is computed as a map of that one state. A change to the state or to a current then moves all projections together, which is how an elastic form factor can inform a transverse-momentum readout. Small classical and quantum geometric nodes represent the residual nonperturbative transverse response, and the GTMD is tied to the collinear sector through the proper soft subtraction, one-loop matching, and evolution~\cite{echevarria,bertone,bertone2025}. The construction language is that of geometric deep learning~\cite{bronstein}: symmetry fixes the admissible transformations, invariant features condition the shared responses, and the physical state metric identifies redundant coordinates.

The proton is the natural first target. A spin-$1/2$ hadron already demands nontrivial orbital, spin, flavor, and color structure, and no other hadron has as many external constraints. We work with the real, T-even, uncharged valence-$u$ $F_{1,1}$ at $\xi=0$, $\mu=2\,\GeV$, $\zeta=4\,\GeV^2$, which has a definite operator definition once the collinear boundary condition and the model for the joint momentum--transfer dependence (the joint closure of Sec.~\ref{sec:physical}) are specified. The inputs are correlated projections of the ART25 TMD extraction~\cite{art25} (ART in what follows), a selection of HERA inclusive data~\cite{hera}, and lattice, current, and form-factor information. Each enters through its own response map with its published covariance, and we keep track of which inputs rest on common underlying data.

The calibration gives two main results. The first is that forward transverse information matters a great deal. At $x=0.4$ and $|\bm k|=0.2\,\GeV$, requiring compatibility with the ART projections narrows the attainable range of $F_{1,1}$ by $81\%$ within the kernel family we use, and by $83\%$ at the nonforward anchor point of Table~\ref{tab:anchors}. The second is that the joint dependence on quark transverse momentum and momentum transfer is a direction none of the present inputs reaches. We write down a deformation, equal to one at $\bm\Delta=0$ and $1+O(r^2)$ at short distance, that leaves every forward and elastic criterion unchanged and moves the nonforward value from $0.588$ to $0.378\,\GeV^{-2}$. The dense momentum-space surfaces and finite-window phase-space maps of Sec.~\ref{sec:tomography} show what that freedom looks like in the images one would like to draw. Together the two results say what present data determine about the proton's five-dimensional structure, and what kind of measurement would be needed to determine the rest.

The architecture was chosen with the next targets in view. In the proton, the symmetry-resolved network gives exact color contraction, exchange statistics that are checked on the reconstructed maps, charge-conserving bonds, and a clean separation between reorganizing a representation, truncating it, and adding physical capacity. In the neutron, the deuteron, and spin-$3/2$ nuclei these same features become essential. Isospin transport becomes a relabeling of operator blocks on the same state; nuclear composition is a tensor product of validated subunits with a one- and two-body operator hierarchy; higher polarization ranks enter through larger symmetry intertwiners whose rank requirements can be measured on the actual response matrices. The proton construction reported here is the first validated step of that program.

A few words are used in specific senses. \emph{Topology} means the organization and connectivity of the finite network and its symmetry-resolved factorizations; Berry geometry, gauge-compatible color contractions, and prescribed holonomies are additional structures with their own tests. Network topology has nothing to do with the topological charge of QCD, and a global color-singlet contraction is not the local Gauss law of a regulated gauge theory. Keeping these apart lets us say what an architectural change does and what it does not. The quantum circuits, likewise, are concrete feature maps and state encodings verified against their exact contractions. They give us a representation whose capacity and cost can be measured; we claim no hardware quantum advantage.

Section~\ref{sec:information} explains how each kind of information enters the shared construction. Section~\ref{sec:architecture} describes the tensor-network state, the geometric nodes, and the validation protocol. Section~\ref{sec:physical} gives the operator definition, matching, evolution, and joint closure that turn the shared state into a physical $F_{1,1}$. Section~\ref{sec:construction-results} states what the calibration establishes, and Sec.~\ref{sec:tomography} presents the momentum-space and phase-space readouts. Section~\ref{sec:extensions} lays out the neutron, deuteron, and spin-$3/2$ constructions, and Sec.~\ref{sec:discussion} concludes. The calibration procedure, source residuals, and conditional ranges are given in full in the appendices so that the numbers quoted in the paper can be traced.

\section{How experimental and phenomenological information enters}
\label{sec:information}
\subsection{A common model with several observable projections}
The organizing object is a common hadronic construction with shared amplitudes, currents, and transverse response functions. Each source of information is compared with the appropriate projection of it. Schematically, with $\theta$ the shared parameters,
\begin{equation}
 y_a^{\rm model}=\mathcal M_a[\theta;\mathcal C_a],
 \qquad
 F_{1,1}^{\rm model}=\mathcal G[\theta;\mathcal C_{\rm GTMD},N].
 \label{eq:input-projections}
\end{equation}
Here $\mathcal M_a$ is the response map for source $a$, $\mathcal C_a$ collects its normalization, operator, and scale conventions, and $N$ is the residual joint closure. The GTMD map $\mathcal G$ combines the common construction with its own matching prescription. Since every projection is evaluated on the same state, a change in a shared amplitude or current propagates to all of them at once. Separate fits to each observable would lose this connection.

Table~\ref{tab:input-roles} summarizes what each class of information does. Symmetry and operator requirements define the admissible structures. Experimental observations and lattice results test particular matrix elements or cross-section maps. Phenomenological extractions contribute model-dependent information on distributions that are not measured directly. The long-distance joint dependence that remains is modeled explicitly, and the last column of the table records, input by input, what is left to that model.

\begin{table*}[t]
\caption{Roles of the information used in the proton construction. Each input enters the shared state through its own response map, together with the restrictions of the original analysis and a record of the data it shares with other inputs. The last column identifies the part of the GTMD that the input leaves to the model; it describes the reach of the input, not the hadron.}
\label{tab:input-roles}
\begin{ruledtabular}
\begin{tabular}{p{0.21\textwidth}p{0.35\textwidth}p{0.35\textwidth}}
Information & Entry into the shared construction & Left to the model\\
\hline
Symmetry and operator structure & Admissible carrier, color and exchange contractions, target helicity, normalization, and matched vector projection & The dynamical state within the admissible family\\
Inclusive scattering: HERA & Conditional DIS response of the common longitudinal and current construction & Intrinsic momentum--transfer correlations; data shared with collinear PDF fits\\
Elastic and static observables & Shared charge, Dirac/Pauli, and transfer-dependent current responses & Intrinsic transverse density; Born, isospin, and nuclear-extraction assumptions\\
Lattice matrix elements & Axial/tensor charges and vector/axial moments through their effective maps & Dependence beyond the finite moments; conditional matching and joint covariance\\
Phenomenological TMD information: ART25 & Correlated forward transverse projections used to calibrate normalized kernels; separately specified evolution ratio & Nonforward joint dependence; extraction assumptions and shared underlying data\\
Explicit nonperturbative closure & Links intrinsic transverse shape to transfer dependence after small-distance matching & Alternatives and deformations invisible to the inputs, quantified in Sec.~\ref{sec:construction-results}
\end{tabular}
\end{ruledtabular}
\end{table*}

\subsection{Incorporation, calibration, and compatibility}
An input is incorporated once both the source quantity and its model prediction are defined: target and flavor assignment, electromagnetic charge factors where they apply, scale and subtraction conventions, and a response map whose approximations are stated. For a phenomenological distribution, we carry along the assumptions of the extraction and the data on which it is based. The ART input, for example, is a correlated ensemble of distribution projections and is treated as one.

The shared model is calibrated against these projections and tested for compatibility. The fit we report uses the full ART covariance on its supported subspace and keeps the criteria for HERA and for the other inputs below fixed caps (Appendix~\ref{sec:inference}). Published uncertainties and nuisance-parameter penalties are used as given; where a covariance is missing, we say so and treat its absence as a conditional assumption rather than as independence. Because several inputs rest on common data and their joint covariance is incomplete, we read the criteria as compatibility diagnostics and not as a global likelihood.

Compatibility in the present fit is approximate. The ART residual is still large, and agreement can trade between source blocks. Appendices~\ref{sec:inference} and \ref{sec:results} record which information is accommodated, which is in tension, and which directions the present sources do not see. That record indicates where the model, the matching, or the inputs should be improved next.

Refinement follows the physics of the response. One identifies the projection a source probes, checks whether the shared architecture contains the required operator or coupling, and tests whether the geometric node can represent the resulting function. A topology change, a new physical amplitude, and additional neural features answer different questions, and each is judged on consistency, representation accuracy, and cost as well as on source compatibility.

\subsection{What this organization enables}
\begin{figure*}[t]
\centering
\includegraphics[width=\textwidth]{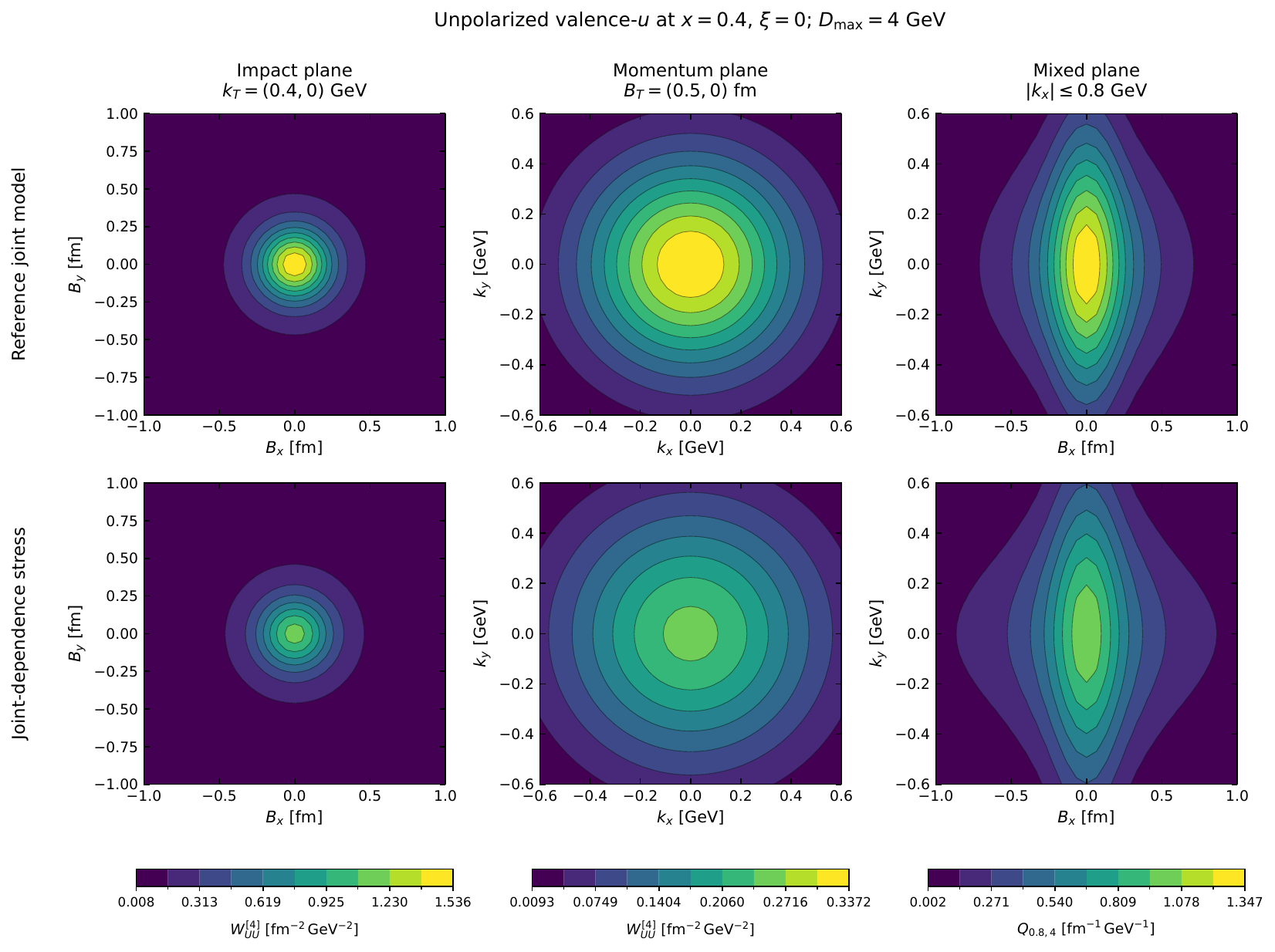}
\caption{Filled-color contour views of the real T-even valence-$u$ model at $x=0.4$, $\xi=0$, $\mu=2\GeV$, and $\zeta=4\GeV^2$. Columns show the impact plane at $\bm k=(0.4,0)\GeV$, the momentum plane at $\bm B=(0.5,0)\,{\rm fm}$, and the mixed $(B_x,k_y)$ projection with $|k_x|\leq0.8\GeV$. Both rows are unpolarized: the upper row uses the reference joint closure, and the lower row applies the illustrative joint-dependence stress of Eq.~\eqref{eq:blind}. They satisfy the same present input criteria. Each column uses a common color range across its two rows. All transforms retain $|\bm\Delta|\leq4\GeV$; the first two columns display $W/(\hbar c)^2$ and the mixed column displays $\mathcal Q/(\hbar c)$, with the distinct units shown. These are fixed-$x$ valence-quark views in the layout of Ref.~\cite{chakrabarti}; that work's $x$-integrated gluon and polarized distributions are different quantities. Transform definitions and domain checks appear in Sec.~\ref{sec:tomography}.}
\label{fig:phase-space-contours}
\end{figure*}

Once the model is fixed, it can be evaluated in GTMD and phase-space variables that were never fitted. Figure~\ref{fig:phase-space-contours} shows the three unpolarized projections used in the upper row of Fig.~4 of Ref.~\cite{chakrabarti}: the impact plane at fixed quark momentum, the momentum plane at fixed impact parameter, and the mixed position--momentum projection. The second row changes the joint-model assumption while holding every present input response fixed. The two rows differ visibly. The images depend on the empirical information and on the continuation into directions that information does not reach, and the construction keeps the two separable.

\section{Shared geometric construction}
\label{sec:architecture}

The architecture combines a finite, symmetry-resolved hadronic representation,
shared effective currents, and classical and quantum geometric function
blocks.  A change of representation, a change of the physical model, and a new
observational constraint are three different operations, and the
construction keeps them apart so that each can be tested on its own.  Geometric deep learning supplies the construction
language: symmetry determines admissible transformations, invariant features
condition shared responses, and the physical state metric identifies
redundant coordinates~\cite{bronstein}.  The neural blocks are small, and each one is checked against an exact
contraction, so what the architecture does is established by direct test.

\subsection{Finite Fock carrier and shared matrix elements}
\label{sec:shared-carrier}

The finite proton construction retains the sectors
\begin{equation}
 \mathcal H_{\mathrm{fin}}=
 \mathcal H_{qqq}\oplus\mathcal H_{qqqg}
 \oplus\mathcal H_{qqqq\bar q}\oplus\mathcal H_{qqqgg}.
 \label{eq:fock-sectors}
\end{equation}
Longitudinal resolution, oscillator support, and operator content are finite model choices, and we state them so that
they can be enlarged in a controlled way.  The dimension of the reduced
coefficient space is a representation count, not a parton count, and
computational qubits are not physical constituents.  A
common coefficient vector $c$ enters the two target-helicity embeddings,
\begin{equation}
 |\Psi_\Lambda(p)\rangle=V_\Lambda(p)c,
 \qquad c^\dagger G_\Lambda c=1,
 \qquad G_\Lambda=V_\Lambda^\dagger V_\Lambda .
 \label{eq:shared-amplitude}
\end{equation}
The embeddings include the finite sector maps and symmetry conventions, and
their dependence on a width or other physical parameter is carried
along whenever that parameter is varied.  Normalization uses the physical
Gram matrix.  The fixed-family geometry below is computed within a
target-helicity fiber; equality of the two helicity Grams is an embedding
check that the 37/38-dimensional construction satisfies within
numerical precision.

For an effective operator, the shared response is
\begin{equation}
 \begin{split}
 W^{[\Gamma]}_{\Lambda'\Lambda}(\kappa)
 &=c^\dagger\mathcal O^{[\Gamma]}_{\Lambda'\Lambda}(\kappa)c,\\
 \mathcal O^{[\Gamma]}_{\Lambda'\Lambda}
 &=V_{\Lambda'}^\dagger(p')\,
 O^{[\Gamma]}(\kappa;\omega)\,V_\Lambda(p),
 \end{split}
 \label{eq:shared-operator}
\end{equation}
where $\omega$ denotes shared current parameters and $\kappa$ the external
kinematics.  Incoming and outgoing amplitudes vary coherently, and current,
moment, and bilocal reductions use the same operator coefficients before
taking their respective projections.  A form factor, a lattice moment, and a transverse projection are thus three
reductions of one bilinear in one state.  At fixed physical operator
parameters, a target spin density matrix enters these effective matrix
elements linearly, and that linearity is what will later carry tensor and
octupole polarization through the same machinery.

Equation~\eqref{eq:shared-operator} has a specific scope.  The final
matched $F_{1,1}$ also uses collinear matching, evolution, and a normalized
joint-shape ansatz; its aggregate shape involves ratios of native bilinears
and is nonlinear in $c$ when the state varies.  The construction parametrizes a correlator with a specified QCD operator
convention; effective operator consistency and physical operator matching
are separate requirements and are tested separately.

\subsection{Physical tensor topology and controlled changes of capacity}
\label{sec:ttn-topology}

The tensor-tree construction exposes collective longitudinal
($L$), Jacobi-orbital ($O$), color-singlet multiplicity ($C$), and joint
spin--flavor--helicity ($S$) registers as physical factor labels.  Fixed
SU(3) singlet intertwiners supply the color contractions, and internal bonds
carry additive doubled $J_z$ and $I_z$ labels, so that each transfer tensor
has the structure
\begin{equation}
 T^{(q,\alpha)}_{(q_1,\alpha_1)(q_2,\alpha_2)}
 =\delta_{q,q_1+q_2}\,
 A^{q;\alpha}_{q_1\alpha_1,q_2\alpha_2},
 \qquad q=(2J_z,2I_z),
 \label{eq:charge-tensor}
\end{equation}
where $\alpha$ labels multiplicity.  This is the implemented additive-charge
version of the separation of symmetry-fixed and degeneracy tensors in
symmetric networks~\cite{singh}.  Color-singlet structure is exact, angular-momentum and isospin projections
are conserved along every bond, and permutation statistics and parity are
checked on the reconstructed physical maps.  None of this is imposed by a
penalty term.  For a composite target, where the spin--flavor--color
register grows and many exchange and symmetry constraints must hold at once,
this is what keeps the construction tractable where a hand-built
parametrization is not.  The global singlets and charge labels are a
kinematic organization; local Gauss's law and a general SU(3) parton-fusion
compiler would be separate constructions.

The physical TTN was validated with 37 complex amplitudes.  Three tree
seeds and one local nonseed reconnection were examined in each of the eight
sector/helicity cells, giving 32 full-rank candidates.  The retained forests
reproduce the state family and the tested current and bilocal matrices with a maximum map discrepancy of $1.82\times10^{-13}$.
Bond dimensions were not selected from discarded singular values alone: full
reconstructions also had to satisfy the exchange and other physical checks,
and two capped forests that failed exchange constraints were rejected.  Those two
forests would have violated Fermi statistics, and the exchange checks excluded them before any fitting was done.

The construction distinguishes three operations on a representation.  Exact
reconnection preserves the state image and changes contraction cost or chart
conditioning.  Rank truncation reduces an image and requires a new
physical-error test.  An independent amplitude adds physical capacity.  The last operation was exercised by adding a 38th complex amplitude orthogonal to the existing 37 columns; it left the original columns unchanged and added two real projective dimensions.  Because the three operations are separately available and separately
tested, enlarging the target from proton to neutron to nucleus can proceed as a sequence of steps that are checked one at a time.

The final fit uses a 114-dimensional carrier, validated subsequently, and a
normalized \emph{native real} state chart with 113 local tangent directions;
its fixed physical embedding may be complex.  This chart is distinct from the
37/38-complex-amplitude prototype.  The physical-register TTN tests on the earlier family apply to that family, and the 114-coordinate chart has its own checks.  Exact
adjacent and root circuit organizations provide alternative encodings of the
same native real family; both attain tangent rank 113 at the initial and final states of the fit, with state-dependent conditioning.  The physical
matching stage changes neither topology nor hidden-node capacity.

\subsection{Classical and quantum geometric nodes}
\label{sec:geometric-nodes}

\begin{figure*}[t]
\centering
\includegraphics[width=\textwidth]{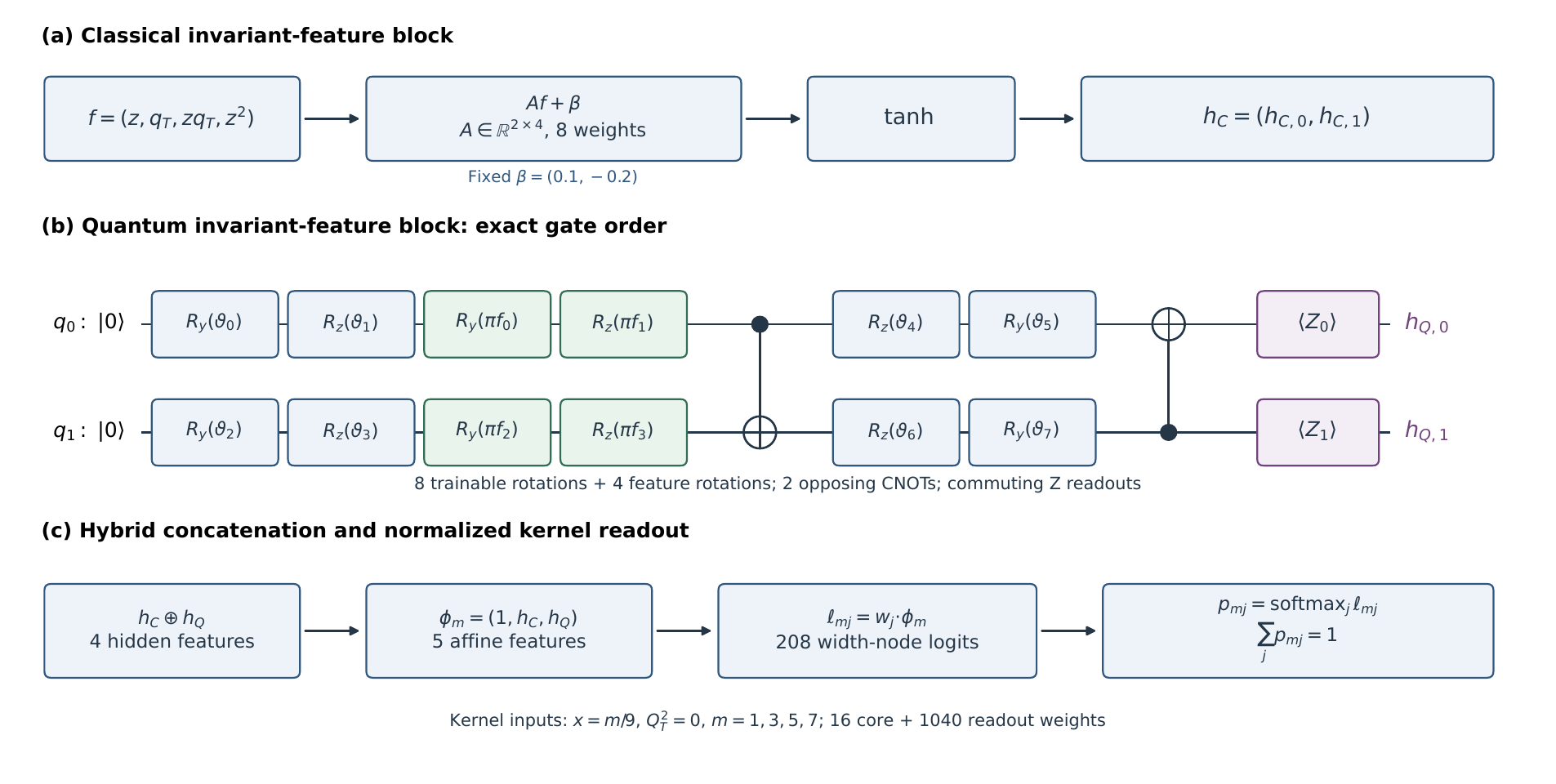}
\caption{Explicit geometric response blocks.  (a) The classical block has
eight trainable entries in $A$, fixed biases, and two $\tanh$ outputs.
(b) The quantum block follows the displayed gate order, with eight trainable
rotations, four feature rotations, two opposing CNOTs, and the commuting
$Z_0,Z_1$ expectations.  (c) Their four outputs are concatenated and augmented
by an affine bias.  The kernel implementation uses 208 logit readouts and a
softmax normalization for each of the four physical modes.  The circuit
is evaluated as a noiseless feature map and checked against its exact
state-vector contraction.}
\label{fig:geometric-nodes}
\end{figure*}

Two node interfaces were constructed.  A state-preparation node acts on known
coherent amplitudes; a response node takes invariant external kinematics or
physical mode labels.  In the 37/38-dimensional prototype, the classical
state node uses
\begin{equation}
 f_c=\bigl(|c_a|^2,|c_b|^2,
             \operatorname{Re}c_a^*c_b,
             \operatorname{Im}c_a^*c_b\bigr).
 \label{eq:coherent-features}
\end{equation}
A fixed three-node $\tanh$ layer feeds eight trainable affine weights, producing
two coherent rotation angles.  The quantum node applies a controlled
two-level rotation and a phase; the hybrid composes noncommuting classical-
conditioned and quantum rotations.  All retain the full amplitudes.  The classical preprocessing assumes amplitude access, so the
state-preparation rule is a nonlinear map on known amplitudes and not a
linear quantum channel on an unknown state.  Six-qubit noiseless circuit checks validate the
specified preparation.  The baseline, classical, quantum, hybrid, and direct
38-dimensional charts have respectively 72, 80, 74, 82, and 74 raw real
coordinates; their generic physical tangent ranks are 72, 74, 74, 74, and 74.
We count physical directions by tangent rank, so redundant neural weights
are never mistaken for physical capacity.

The response nodes use dimensionless scalar features
\begin{equation}
 \begin{split}
 z&=2x-1,\qquad
 q_T=\frac{Q_T^2}{s_0+Q_T^2},\qquad s_0=0.16\;\mathrm{GeV}^2,\\
 f&=(z,q_T,zq_T,z^2),\qquad Q_T^2=-t\quad(\xi=0).
 \end{split}
 \label{eq:probe-features}
\end{equation}
Figure~\ref{fig:geometric-nodes} shows the implemented blocks and their
hybrid readout.
$Q_T^2$ denotes transverse transfer, not the hard DIS scale.  The classical
block is
\begin{equation}
 h_C(f)=\tanh(Af+\beta),\qquad
 A\in\mathbb R^{2\times4},\quad\beta=(0.1,-0.2),
 \label{eq:classical-node}
\end{equation}
with eight trainable entries in $A$.  The quantum block uses two qubits and
eight trainable angles, returning
\begin{equation}
 (h_Q)_a=\langle00|U^\dagger(f,\vartheta)Z_a
 U(f,\vartheta)|00\rangle,\qquad a=0,1.
 \label{eq:quantum-node}
\end{equation}
For each wire $a$, the gate order is $R_y(\vartheta_{2a})$,
$R_z(\vartheta_{2a+1})$, $R_y(\pi f_{2a})$, and
$R_z(\pi f_{2a+1})$; a CNOT from wire 0 to 1 follows.
The second layer applies $R_z(\vartheta_{4+2a})$ then
$R_y(\vartheta_{5+2a})$ on each wire, followed by a CNOT from 1 to 0.
Thus the block has twelve single-qubit rotations, two CNOTs, and two
commuting $Z$ readouts.  Circuit expectations and their exact state-vector contractions define the
same feature family, which is why the quantum block can be verified exactly.
It is a small concrete instance of quantum circuit learning~\cite{mitarai}.

For an operator basis $O_b$, scalar readouts define
\begin{equation}
 O(\kappa)=O_0(\kappa)+\sum_b a_b(\kappa)O_b(\kappa),
 \qquad a_b=w_b\cdot(1,h),
 \label{eq:operator-node}
\end{equation}
where $h$ is $h_C$, $h_Q$, or their concatenation.  Charge, crossing, and tensor conventions live in the basis; the bounded
hidden features only modulate coefficients within it.  A nested extension retains
$(1,z,q_T,zq_T,z^2)$ and appends blocks; setting the added readout weights to
zero recovers the parent exactly, which gives a controlled capacity
comparison.  The final physical fit uses six shared current coordinates,
including the Dirac transfer response $D=1+dq_T+eq_T^2$, and a transverse
width.  This current-family extension adds explicit response freedom without
adding a Fock direction or a neural hidden unit.

\subsection{Normalized transverse kernels and finite-mode expressivity}
\label{sec:kernel-network}

The response node used in the final model supplies four normalized
transverse kernels,
\begin{equation}
 \begin{split}
 K_m(r)&=\sum_{j=1}^{208}p_{mj}
                 \exp(-\eta_j r^2/4),\quad m\in\{1,3,5,7\},\\
 p_{mj}&=\frac{\exp\ell_{mj}}{\sum_{k=1}^{208}\exp\ell_{mk}},
 \qquad\ell_{mj}=w_j\cdot\phi_m,\\
 \phi_m&=(1,h(f(x=m/9,Q_T^2=0))).
 \end{split}
 \label{eq:geometric-kernel}
\end{equation}
Here $r$ is the transverse separation conjugate to parton momentum, and
$\eta_j\in[0,4]\;\mathrm{GeV}^2$ is a fixed finite grid.
Each probability row has unit mass and $K_m(0)=1$, so the collinear boundary, which is fixed separately, is preserved; evolution and soft/rapidity matching
multiply outside this node.  The nonnegative Gaussian-mixture weights
restrict the kernel family.  They say nothing about the sign of the
renormalized nonforward correlator, which the matched construction leaves
free.

\begin{table}[t]
\caption{Implemented kernel-node sizes.  Ranks include the affine bias and
refer to the four physical mode inputs at the checked cores.  Raw weights
include redundant directions; the rank is the physically meaningful count.}
\label{tab:kernel-node-sizes}
\centering
\begin{tabular}{lrrrrr}
\hline\hline
Block & Core & Hidden & Readout & Total & Rank\\
\hline
Classical & 8 & 2 & 624 & 632 & 3\\
Quantum & 8 & 2 & 624 & 632 & 3\\
Hybrid & 16 & 4 & 1040 & 1056 & 4\\
\hline\hline
\end{tabular}
\end{table}

The complete four-row probability table has $4(208-1)=828$ independent
coordinates.  If the $4\times5$ hybrid design matrix $\Phi$ has row rank four,
any strictly positive table can be realized at fixed core: choose
row-centered target logits and solve $\Phi W^{\mathsf T}=\log p$ up to the
irrelevant row constants.  This rank argument is how we size the node.  It shows that the hybrid block
is sufficient for the four physical modes of the proton, and that the
capacity required is set by the number of physical modes rather than by the
number of hidden units.  We will apply the same rule when a target with more
modes or a higher polarization rank enlarges the response matrix.  The
argument concerns the finite set of physical mode inputs; it is not a
universal approximation statement over continuous kinematics.

The final fitted table has 113 exact zero entries, which finite softmax logits
cannot reproduce.  The realized table therefore uses the explicit interior
replacement $p\mapsto(1-\epsilon)p+\epsilon/208$ with
$\epsilon=10^{-12}$.  Its maximum probability displacement is
$9.95\times10^{-13}$.  The literal two-qubit circuit reproduces the analytic
features to $3.33\times10^{-16}$, and its ART criterion changes by
$2.55\times10^{-7}$ from the exact rational reference; all quoted results use the exact normalized fitted table.  This final realization fixes the core and solves for the readout weights.
It is an interpolation check, separate from the earlier neural training
comparisons, and it shows that the quantum--classical node reproduces the calibrated response.

\subsection{State geometry, gauge freedom, and observable sensitivity}
\label{sec:state-geometry}

For normalized states with real parameters $\theta$, the quantum geometric
tensor and its metric are
\begin{equation}
 \begin{split}
 Q_{ab}&=\langle\partial_a\Psi|
 (1-|\Psi\rangle\langle\Psi|)|\partial_b\Psi\rangle,\\
 g_{ab}&=\operatorname{Re}Q_{ab}.
 \end{split}
 \label{eq:qgt-architecture}
\end{equation}
With $\mathcal A_a=i\langle\Psi|\partial_a\Psi\rangle$ the curvature convention
is $\mathcal F_{ab}=-2\operatorname{Im}Q_{ab}$.
For unnormalized coefficients, fixed Gram $G$, derivative columns
$J_a=\partial_a c$, and $N=c^\dagger Gc$, the implemented pullback is
\begin{equation}
 Q_{ab}=\frac{J_a^\dagger GJ_b}{N}
 -\frac{(J_a^\dagger Gc)(c^\dagger GJ_b)}{N^2}.
 \label{eq:gram-qgt}
\end{equation}
Parameter-dependent physical maps require the full derivative of $Vc$.  The
Gram is contracted through nested TTN transfers and physical root columns,
providing a check independent of the coefficient chart.

A compensated bond change $T\mapsto T U$, with $U$ restricted to equal-charge
blocks and $U^\dagger$ on the parent, leaves the physical state unchanged.
Global phase and such bond gauges are removed before counting rank, and
purification redundancy is separated from a physical target mixture.
Berry-loop tests measure transport within the finite family; the
tested ordered SU(3) transport binds that family to an independently
prescribed classical connection.

The state metric is distinct from the sensitivity of a source prediction
$y(\theta)$.  For its full supported covariance $C$,
\begin{equation}
 (g_{\mathrm{obs}})_{ab}
 = (\partial_a y)^{\mathsf T}C^+(\partial_b y),
 \label{eq:observable-pullback}
\end{equation}
where $C^+$ inverts only the supported covariance subspace.  Comparing $g$ with $g_{\rm obs}$ is a useful diagnostic.  A direction with nonzero $g$ and zero $g_{\mathrm{obs}}$
is one the representation resolves and the current observations do not.
The comparison tells us whether a failed fit calls for a different solver, an
extended physical response, or new input, and it is how the joint-dependence
freedom of Sec.~\ref{sec:construction-results} was first noticed.  A natural-gradient
update based on the supported $g^+$ is an available geometric optimization scheme~\cite{stokes}.  The reported kernel-node training used analytic
derivatives with L-BFGS-B, and the final physical result used direct
constrained state/current/width fitting and convex kernel fitting.

\subsection{Validation and refinement protocol}
\label{sec:architecture-validation}

Each refinement starts from a fixed physical question and from a parent model that the extended model reproduces exactly when the added parameters are set to zero. The state family, operator definitions, input definitions, and numerical tolerances are fixed before a new variant is tested.  Structural validation
checks both target helicities, exchange statistics, normalization, current
identities, and both operator legs.  Representation validation compares
classical contractions with literal circuits and compares analytic
derivatives with independent differences or gate-shift derivatives.
Geometry validation checks the physical Gram, quotient rank, compensated
bond gauges, and Berry conventions.  A fit improvement is then assessed
with source-block residuals, model assumptions, and total execution cost
recorded separately.

We applied this protocol at each stage.  The five early amplitude-node
configurations passed state, circuit, Jacobian, QGT, and readout checks.
Later classical, quantum, and hybrid kernel fits used two starts each and were
compared with the complete fixed-state simplex.  Since the hybrid node had sufficient finite-mode rank and the direct kernel
fit succeeded, the final calibrated result was loaded into the existing node;
the capacity needed was known from the rank analysis before any additional
hidden units came into question.  Throughout, we distinguish between an optimizer reporting convergence, an attained improvement of the criterion, constrained stationarity, and global optimality.

Costs are recorded alongside capacity.  In the early validated forest,
fixed-map buffers fell from 43.2 to 18.5 MB, while median state synthesis
rose from 3.9 to 16.6 ms.  The adjacent native 114-dimensional
circuit uses 114 queued multicontrolled-X operations versus 388 in the root
organization; both use 113 controlled $R_y$ rotations and the same dense
seven-qubit encoding.  These are logical counts before hardware decomposition; a complete quantum
cost would also include amplitude access, feature evaluation, parameter
loading, synthesis, measurement, and readout.  What we have at this stage are
faithful, reproducible classical and quantum realizations with measured
capacity and cost.  The comparison between them becomes interesting on the
larger response matrices of nuclear targets, and that is where we intend to
make it.

\section{A physically matched \texorpdfstring{$F_{1,1}$}{F1,1} construction}
\label{sec:physical}
\subsection{Operator and projection}
We use the metric $(+,-,-,-)$ and light-front components $v^+=(v^0+v^3)/\sqrt{2}$. The lightlike vector $n$ has $n^+=0$, $n^-=1$, and $\bm n_T=0$, so $P\cdot n=P^+$. At zero skewness, $p^+=p'^+=P^+$, $\bm P_T=0$, $\bm p_T=-\bm\Delta/2$, $\bm p'_T=\bm\Delta/2$, and $t=-\bm\Delta^2$. The transverse separation $\bm r$ is conjugate to quark transverse momentum $\bm k$. It must be distinguished from the impact parameter $\bm B$, which is conjugate to transfer $\bm\Delta$.

Our vector helicity correlator is the UV-renormalized, soft-subtracted matrix element of
\begin{equation}
 \bar q\bigl((zn+r_T)/2\bigr)\,
 U_s\,\frac{\gamma^+}{2}\,
 q\bigl(-(zn+r_T)/2\bigr),
\end{equation}
Fourier transformed with $\int \dd z\,e^{-\ii xP^+z}/(2\pi)$ and divided once by $\sqrt{S_q}$. Here $U_s$ is a fundamental staple with orientation $s=\pm1$, and $S_q$ is the corresponding soft factor. The finite subtraction and causal prescription follow the matched convention of Ref.~\cite{bertone2025}; a straight Wilson line is not substituted. The scales $\mu$ and $\zeta$ have units of energy and energy squared, respectively. The real T-even scalar is obtained by averaging the scalar coefficient over the two staple orientations, rather than by taking the entrywise real part of the helicity matrix.

With $\bar u u=2M$, the unpolarized helicity trace selects
\begin{equation}
 \Ft=\frac12\Tr \widetilde W^{[\gamma^+]}_{u-\bar u},
 \qquad
 \F=\int\frac{\dd^2\bm r}{(2\pi)^2}
 e^{+\ii\bm k\cdot\bm r}\Ft .
 \label{eq:fourier}
\end{equation}
The trace identity holds at $\xi=0$, including the forward and collinear degeneracies. It does not determine companion amplitudes that become invisible in those limits. The positive-$x$ antiquark convention is
\begin{equation}
 \widetilde W_{\bar q}(x,\bm r,\bm\Delta)
 =-\widetilde W_q(-x,-\bm r,\bm\Delta).
\end{equation}
The target is uncharged: no electromagnetic charge factor or additional factor of $x$ is included. With dimensionless collinear GPD conventions, $\Ft$ is dimensionless and $\F$ has units $\GeV^{-2}$ per unit $x$.

At leading order in the short-distance expansion, the vector matrix contains both matched $H$ and $E$. Its half trace selects the $H$ term at $\xi=0$, while $E$ remains in the spin-flip matrix of the same construction. Similarly, the one-loop gluon coefficients cancel between a flavor and its antiquark in the nonsinglet difference, while gluon and sea degrees of freedom remain in the shared proton carrier for the singlet and polarized projections.

\subsection{Short-distance matching and scales}
For $r\ll\Lambda_{\rm QCD}^{-1}$, the nonsinglet result is
\begin{equation}
 \begin{split}
 \Ft(x,\bm r,\bm\Delta;\mu,\zeta)
 ={}& [C_{\rm NS}\otimes H_v](x,t;\mu,\zeta,r)\\
 &+\text{power corrections},
 \end{split}
 \label{eq:ope}
\end{equation}
where $(C\otimes H)(x)=\int_x^1\dd y\,C(y)H(x/y)/y$.
Writing $a_s=\alpha_s/(4\pi)$, $C_F=4/3$, $b_0=2e^{-\gamma_E}$, $L=\ln(\mu^2r^2/b_0^2)$, and $\ell=\ln(\mu^2/\zeta)$, the complete one-loop coefficient used here is~\cite{bertone,bertone2025}
\begin{align}
 C_{\rm NS}(y)={}&\delta(1-y)+a_s C_F\Bigg[
 -2L(1+y^2)\left[\frac{1}{1-y}\right]_+
 \nonumber\\[-2pt]
 &+2(1-y)+\delta(1-y)
 \left(-L^2+2L\ell-\frac{\pi^2}{6}\right)\Bigg].
 \label{eq:coefficient}
\end{align}
The plus prescription acts on the denominator only. At canonical scales, $L=\ell=0$, the finite regular and delta terms remain. Dropping them would replace the matched response by a different approximation. Appendix~\ref{app:plus} specifies the endpoint implementation.

The model identifies its effective shared $H/E$ boundary with an $\MS$ collinear input at $\mu_0=2\GeV$. This identification is a stated model assumption, and it is the point at which a measured off-forward boundary from exclusive data can later replace the assumed one. For either GPD the longitudinal reconstruction has the form
\begin{equation}
 \begin{split}
 H_v(x,t;\mu_0)&=\sum_{m\in\{1,3,5,7\}}B_m(x)H_m(t),\\
 B_m(x)&=\frac{x^{m-1}(1-x)^{8-m}}{\mathrm B(m,9-m)}.
 \end{split}
 \label{eq:beta}
\end{equation}
Each $B_m$ integrates to one and has mean $m/9$. These functions reconstruct a continuous distribution from four longitudinal modes; their higher moments are model information.

We evolve the nonsinglet modes with HOPPET~\cite{hoppet} at NLO, using $\alpha_s(M_Z)=0.118$ and thresholds $1.4$, $4.5$, and $175\GeV$. The evolution uses an $x$-grid parameter $dy=0.00625$, checked against $dy=0.008$. The matching scale is regulated by
\begin{equation}
 r_* =\frac{r}{\sqrt{1+(\mu_{\rm floor}r/b_0)^2}},
 \qquad \mu_i=\frac{b_0}{r_*},\qquad \zeta_i=\mu_i^2.
 \label{eq:rstar}
\end{equation}
The reference floor is $2\GeV$; $1.5$ and $3\GeV$ are used as variations. The coefficient is evaluated at the canonical separation $r_*$ with $(\mu_i,\zeta_i)$, so its logarithms vanish while its finite one-loop terms remain. The nonperturbative shape and source evolution retain the physical separation $r$. All displayed physical outputs are evolved to $\mu=2\GeV$, $\zeta=4\GeV^2$.

The evolution factor $R_{\rm src}(r)$ is the ART25 central member's ratio of evolved distributions at the same physical $r$, between $(\mu_i,\zeta_i)$ and $(2\GeV,4\GeV^2)$. The intrinsic source distribution cancels in this ratio. Its nonperturbative Collins--Soper kernel is inherited from the source. The source coupling order and its bottom threshold differ from those of the NLO collinear evolution, so the present construction is of mixed perturbative order; replacing $R_{\rm src}$ by a native momentum-space evolution operator at a single perturbative order is a straightforward improvement within the same construction.

For the same perturbative scales, the conversion between our convention and the forward ART convention has no finite nonsinglet term through one-loop order. This does not identify the two complete finite-$r$ ansatzes: ART has its own short-distance transition, higher orders, and flavor-dependent intrinsic shapes. In particular, finite-$r$ quark-minus-antiquark combinations can retain a gluon coefficient multiplying the difference of those shapes. We use the nonsinglet short-distance coefficient and source evolution ratio and fit our own intrinsic remainder.

\subsection{Joint closure and normalized transverse kernels}
The aggregate construction is
\begin{equation}
 \Ft=R_{\rm src}(r)[C_{\rm NS}\otimes H_v](x,t;\mu_i,\zeta_i)
 N(x,\bm r,\bm\Delta).
 \label{eq:model}
\end{equation}
Let $S_m(\bm r,\bm\Delta)$ denote the effective shared overlap in mode $m$, with $S_m(0,\bm\Delta)=H_m(t)$. The reference, or native-joint, closure is
\begin{equation}
 N_{\rm joint}=
 \frac{\sum_m B_m(x)S_m(\bm r,\bm\Delta)K_m(r)}
 {H_v(x,t;\mu_0)}.
 \label{eq:native}
\end{equation}
This ratio is used only where its denominator check passes. A Fourier integral additionally requires an admissible continuous transfer interval: individually safe quadrature nodes can straddle a pole. No denominator floor is inserted. The alternate forward-shape closure sets
\begin{equation}
 N_{\rm fwd}(x,\bm r,\bm\Delta)=N_{\rm joint}(x,\bm r,0),
 \label{eq:forwardshape}
\end{equation}
while retaining $t$ in the matched $H_v$ factor in Eq.~\eqref{eq:model}. Both closures approach one at small $r$ on the validated domain. They agree in the forward input maps and differ in their nonforward predictions.

The four radial kernels are
\begin{equation}
 \begin{split}
 K_m(r)&=\sum_{j=1}^{208}p_{mj}e^{-\eta_jr^2/4},\\
 p_{mj}&\geq0,\quad\sum_jp_{mj}=1,\quad
 0\leq\eta_j\leq4\GeV^2.
 \end{split}
 \label{eq:kernel}
\end{equation}
The Gaussian-mixture family and its width support are restrictions on the response. The finite $\eta$ grid is a further numerical restriction. These positive coefficients parametrize the kernel family; the renormalized, signed, nonforward GTMD built from them is unconstrained in sign.

Because Eq.~\eqref{eq:native} divides one state bilinear by another, the aggregate response is nonlinear in those bilinears as the state changes. A separately tested modewise alternative,
\begin{equation}
 \left.\Ft\right|_{\rm mode}=R_{\rm src}\sum_m(C_{\rm NS}\otimes B_m)(x)
 S_m(\bm r,\bm\Delta)K_m(r),
 \label{eq:modewise}
\end{equation}
removes the aggregate denominator. At fixed width, kernels, and physical current coefficients it is bilinear in native amplitudes, which restores a single state-independent effective operator on the finite carrier. It alters the forward source predictions and therefore requires its own calibration; its tests are reported separately from the final aggregate fit, and Sec.~\ref{sec:tomography} identifies the reason it is the preferred closure for the next iteration.

The residual freedom that the present inputs leave open can be exhibited exactly. Consider
\begin{equation}
 N\longmapsto N\exp\left[-\frac{\eta_Jr^2}{4}
 \frac{\bm\Delta^2}{0.16\GeV^2+\bm\Delta^2}\right].
 \label{eq:blind}
\end{equation}
It equals unity at $\bm\Delta=0$ and is $1+O(r^2)$ at short distance. It therefore preserves every current forward input and the leading matching while changing the joint dependence, and the collinear maps are untouched. We use the fixed illustrative value $\eta_J=0.25\GeV^2$ as a stress test. Composing this factor with Eq.~\eqref{eq:kernel} can exceed the profiled width support, so the stress probes a direction outside the finite family covered by the kernel profiles; this is what makes it informative.

\section{What the calibration establishes}
\label{sec:construction-results}
The construction gives an evaluable model of the physical $F_{1,1}$, with common amplitudes and currents connecting the maps for the individual inputs. It combines operator matching, an explicit long-distance completion, calibration to complementary information, and geometric representations of the resulting response. Since the model is a function of the shared state and not a table of fitted points, it can be inspected in variables that none of the sources constrain directly, with the assumptions visible at each step. We draw two conclusions from the calibration.

\paragraph*{Forward transverse information narrows the GTMD substantially.} At fixed state and prescription, the ART compatibility condition restricts the four normalized kernels and with them the attainable range of $F_{1,1}$. At $x=0.4$ and $|\bm k|=0.2\GeV$ the reference forward value is $0.978\GeV^{-2}$, with an attained inner finite-kernel range of $[0.813,0.997]\GeV^{-2}$; without the ART cap the same kernel family spans $[0.060,1.033]\GeV^{-2}$. The attained width shrinks by $81\%$ forward and by $83\%$ at the nonforward anchor (Table~\ref{tab:anchors}, Fig.~\ref{fig:profiles}). The ranges are conditional on the state and the kernel family, and within that family they measure how much the input matters. Appendix~\ref{sec:results} has the full comparison.

\paragraph*{Joint momentum--transfer dependence is left to the model, and we can say by how much.} The reference native-joint closure and the forward-shape alternative satisfy identical source criteria and give different nonforward readouts. The factor in Eq.~\eqref{eq:blind} is sharper. It preserves every present input and the leading small-distance matching, yet at the nonforward perpendicular anchor it moves $F_{1,1}$ from $0.588$ to $0.378\GeV^{-2}$, a $36\%$ change in a quantity whose conditional range under the present inputs is only $[0.499,0.599]\GeV^{-2}$. Figure~\ref{fig:phase-space-contours} shows the same thing in phase space. Present data therefore pin down the forward transverse structure rather well and the correlation between quark momentum and momentum transfer hardly at all, and we have an explicit deformation that any future constraint would have to respond to.

Calibration improves the agreement of the common model with its inputs: the ART criterion falls by $31\%$ from the initial shared state, though a substantial ART residual remains. The source-block trade-offs, matching-floor sensitivity, and conditional covariance assumptions are kept in the appendices, where they serve as a map of what to improve next. The quantum--classical representation validated earlier reproduces the fitted response functions, so no new topology or hidden-node capacity was needed here; the rank analysis of Sec.~\ref{sec:kernel-network} tells us in advance when it will be.

The next section displays the model's momentum and phase-space structure in detail. All of the views are evaluations of the same fitted model, with explicit transfer and momentum windows where they are needed, and they separate the source-constrained directions from the model-dependent continuation.

\section{Momentum and phase-space readouts}
\label{sec:tomography}
\subsection{Dense readouts at fixed longitudinal fractions}
The fitted construction is an evaluable model, so its analytic transverse structure and the tabulated evolution at $x=0.2,0.4,0.6$ give dense momentum readouts with no new fit. We do not interpolate in $x$; the three longitudinal slices are evaluated independently. Dense evaluation tests the numerical implementation of the fixed model, and the empirical restrictions remain those of Sec.~\ref{sec:profiles}.

The evaluator contracts the four 208-point kernel rows before performing the transform, reuses the exact finite angular harmonics, and reproduces all 36 values computed during the fit (Appendix~\ref{sec:results}). Independent contractions of the individual Gaussian components, simultaneous rotations, transfer-reversal Hermiticity, and an anisotropic Gaussian transform check test signs and normalization. The displayed momentum surfaces are the real T-even valence-$u$ component at $\xi=0$, $\mu=2\GeV$, $\zeta=4\GeV^2$.

\begin{figure*}[t]
\centering
\includegraphics[width=\textwidth]{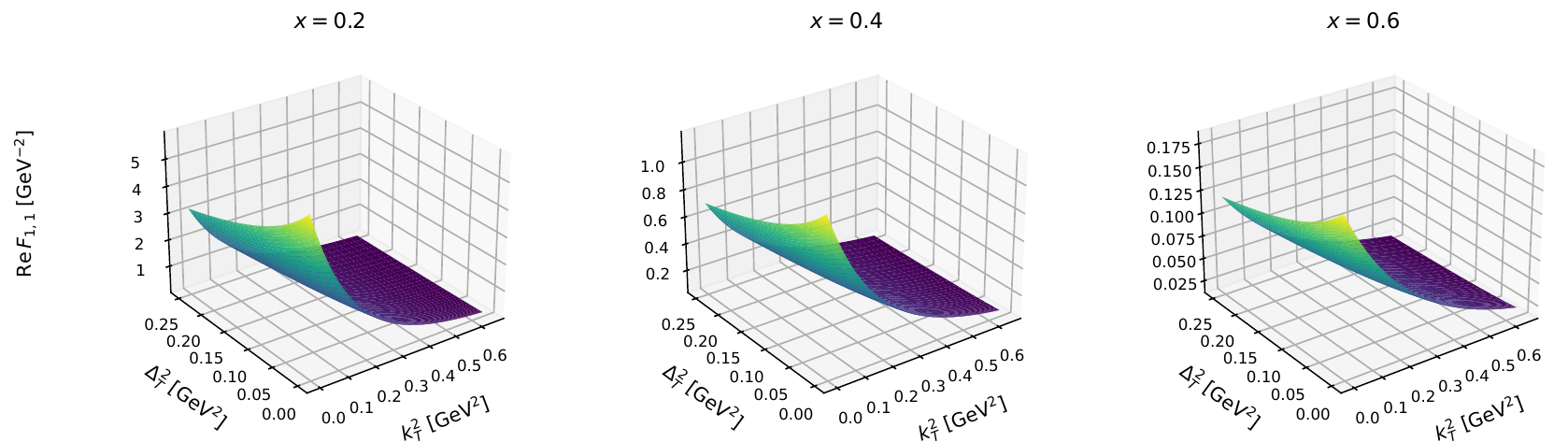}
\caption{Dense readouts of the native-joint $F_{1,1}^{u-\bar u}$ model at $x=0.2,0.4,0.6$, with $\bm k\perp\bm\Delta$. The axes are $k_T^2$ and $\Delta_T^2$; $|\bm k|\leq0.8\GeV$ and $|\bm\Delta|\leq0.5\GeV$. All panels use $\xi=0$, $\mu=2\GeV$, $\zeta=4\GeV^2$. Vertical scales differ to resolve each slice. Dense evaluation visualizes the fitted model; it adds no new information.}
\label{fig:f11-surfaces}
\end{figure*}

\begin{figure*}[t]
\centering
\includegraphics[width=\textwidth]{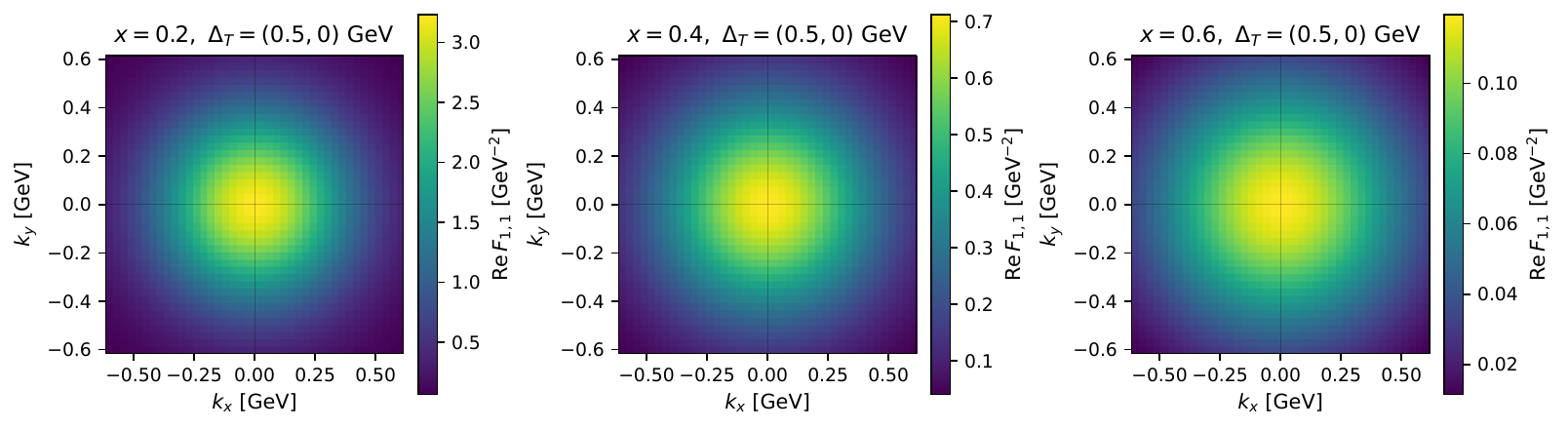}
\caption{Transverse momentum planes of the same model at $\bm\Delta=(0.5,0)\GeV$ and three fixed $x$ values. Each square extends to $|\bm k|=0.849\GeV$ at its corners. Color scales are separate numerical densities in $\GeV^{-2}$; no normalization to unit peak or scalar positivity constraint is applied.}
\label{fig:f11-planes}
\end{figure*}

Figures~\ref{fig:f11-surfaces} and \ref{fig:f11-planes} resolve the momentum and transfer dependence separately from the pointwise profiles. Their changing vertical and color scales are essential: the distributions are per unit $x$, and they are not independently normalized images. The squared momentum axes in Fig.~\ref{fig:f11-surfaces} change only the coordinates used to display $F_{1,1}$, not its density convention.

\subsection{Transfer-domain validation and finite-window readouts}
Extending the transform domain requires a new validation of the model, and carrying it out revealed a concrete limitation. In the native-joint closure, the normalization $H_v(x,-D^2;\mu_0)$ has zeros at approximately
\begin{equation}
 \begin{split}
 D&=6.74,\ 5.70,\ 5.03\GeV\\
 &\hspace{15mm}\text{for }x=0.2,\ 0.4,\ 0.6,
 \end{split}
 \label{eq:poles}
\end{equation}
respectively. The matched numerator stays nonzero at the tested intrinsic separations, and direct near-root evaluations show sign-changing singular behavior, so these are genuine poles of the aggregate continuation. A first sparse survey found safe isolated nodes on either side of them; the interval and root checks of the validation protocol exposed them. The poles lie outside the $D\leq0.5\GeV$ domain of the fitted targets and leave those results unchanged. They also single out the modewise closure of Eq.~\eqref{eq:modewise}, which has no aggregate denominator, as the right continuation to use for an unrestricted Wigner transform.

For the present model we define a finite-transfer-window readout,
\begin{equation}
 W_w(x,\bm k,\bm B)=
 \int\frac{\dd^2\bm\Delta}{(2\pi)^2}
 e^{-\ii\bm\Delta\cdot\bm B}
 w(D)\F(x,\bm k,\bm\Delta).
 \label{eq:window-wigner}
\end{equation}
The reference window is $w_4(D)=\Theta(4\GeV-D)$. Lower hard cutoffs at $2$ and $3\GeV$ and a cosine taper from $3$ to $4\GeV$ test window sensitivity. The latter equals one below $3\GeV$, $\{1+\cos[\pi(D/\GeV-3)]\}/2$ between $3$ and $4\GeV$, and zero above. Each defines a resolved readout of the same model; the window is part of the definition.

The extension from $0.5$ to $4\GeV$ is a tested model extrapolation, not a new empirical constraint, and the window dependence measured below is the resolution statement that goes with the images. A Wigner function is a quasi-distribution, so signed structure in the maps is expected.

\begin{figure*}[t]
\centering
\includegraphics[width=\textwidth]{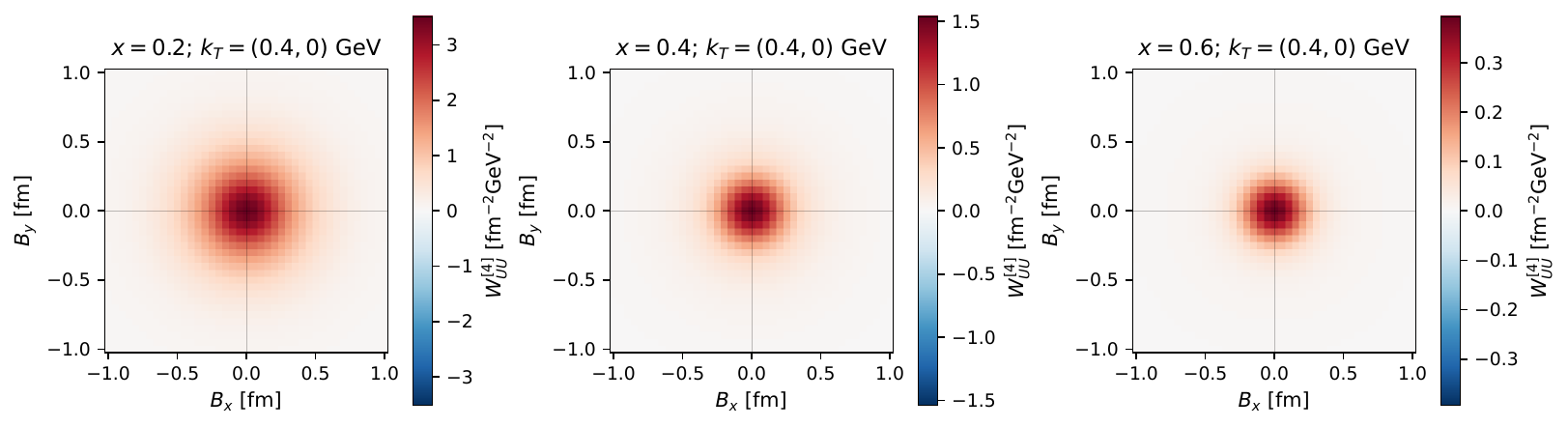}
\caption{Unpolarized finite-transfer-window phase-space maps $W_{UU}^{[4]}$ at fixed $x$ and $\bm k=(0.4,0)\GeV$, using the native-joint closure and the hard circular window $|\bm\Delta|\leq4\GeV$. Here $W_{UU}^{[4]}$ denotes $W_{w_4}^{\rm fm}$. $B_x,B_y$ are impact coordinates conjugate to transfer. The color scale includes $1/(\hbar c)^2$ and has units $\mathrm{fm}^{-2}\GeV^{-2}$. The extension beyond the original transfer anchors is model extrapolation; its window dependence is displayed in Fig.~\ref{fig:window-sensitivity}.}
\label{fig:wigner-B}
\end{figure*}

\begin{figure*}[t]
\centering
\includegraphics[width=\textwidth]{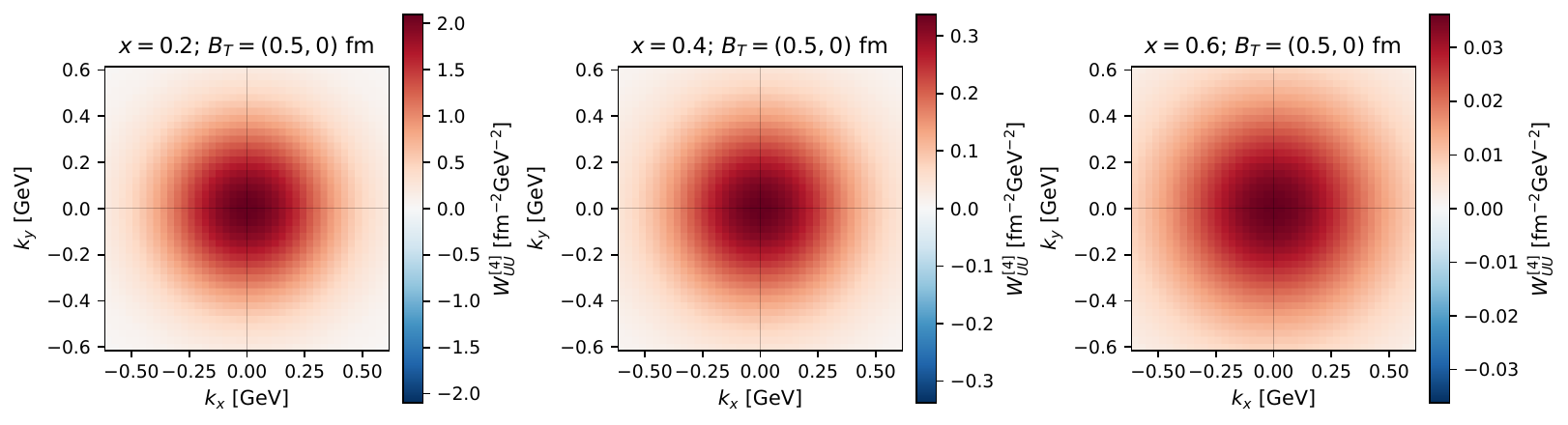}
\caption{Momentum-plane cuts of the same finite-window $W_{UU}^{[4]}$ at $\bm B=(0.5,0)\,{\rm fm}$. The three fixed-$x$ panels use the same transfer window and physical scales as Fig.~\ref{fig:wigner-B}. The square corners reach $|\bm k|=0.849\GeV$. Separate color scales show the actual model densities, including the position-unit Jacobian. These are fixed-$x$ valence-quark maps.}
\label{fig:wigner-k}
\end{figure*}

\subsection{Connection to phase-space visualizations}
The impact-plane map fixes $\bm k$ and varies $\bm B$; the momentum-plane map fixes $\bm B$ and varies $\bm k$. These are distinct cuts of Eq.~\eqref{eq:window-wigner}, distinct from the intrinsic-separation kernel $K_m(r)$. Their density conversion for axes in fm is specified in Appendix~\ref{app:transforms}; multiplication by $(\hbar c)^2$ recovers the natural-unit density.

These unpolarized fixed-$x$ readouts are the present analogues of the $UU$ panels in Ref.~\cite{chakrabarti}. Polarized $UL$, $LU$, or $LL$ maps, and the orbital-angular-momentum observables they carry, need the additional scalar amplitudes of the same construction with their own projectors. They are the natural next readouts once the spin-flip sector has been calibrated.

Figures~\ref{fig:wigner-B} and \ref{fig:wigner-k} illustrate the two kinds of cut. The impact maps use $|B_x|,|B_y|\leq1\,{\rm fm}$, so corner points reach $|\bm B|=\sqrt2\,{\rm fm}$. The numerical tolerances are applied to the complete square grids rather than only their axes.

\subsection{Window and joint-model sensitivity}
\begin{figure*}[t]
\centering
\includegraphics[width=0.85\textwidth]{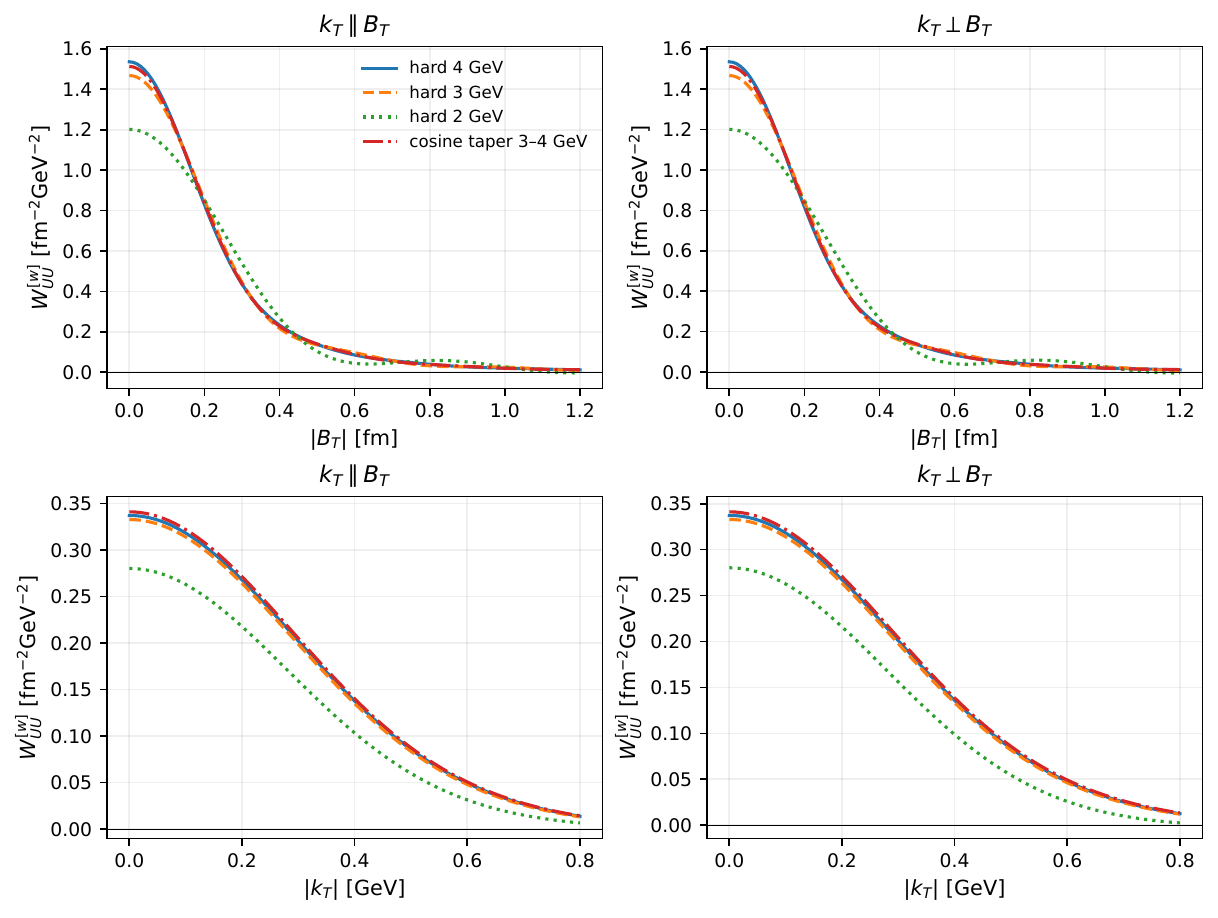}
\caption{Finite-window sensitivity at $x=0.4$. Upper panels fix $|\bm k|=0.4\GeV$ and vary impact distance; lower panels fix $|\bm B|=0.5\,{\rm fm}$ and vary momentum. Left and right columns compare parallel and perpendicular orientations. Hard cutoffs at $2$, $3$, and $4\GeV$ and a cosine taper from $3$ to $4\GeV$ define distinct readouts. Their difference measures resolution-prescription sensitivity.}
\label{fig:window-sensitivity}
\end{figure*}

The finite-window checks have a clear quantitative effect. At $x=0.4$, the hard-$4\GeV$ impact map has a maximum of $1.536\,{\rm fm}^{-2}\GeV^{-2}$. Relative to that map, the largest absolute changes over its displayed plane are $0.334$, $0.068$, and $0.023\,{\rm fm}^{-2}\GeV^{-2}$ for the hard-$2\GeV$, hard-$3\GeV$, and cosine-tapered windows, respectively. These are roughly $22\%$, $4\%$, and $1.5\%$ of the reference peak. As Fig.~\ref{fig:window-sensitivity} shows, the differences are far larger than the integration errors, so the window is the dominant resolution choice and has to be stated with any image.

\begin{figure*}[t]
\centering
\includegraphics[width=\textwidth]{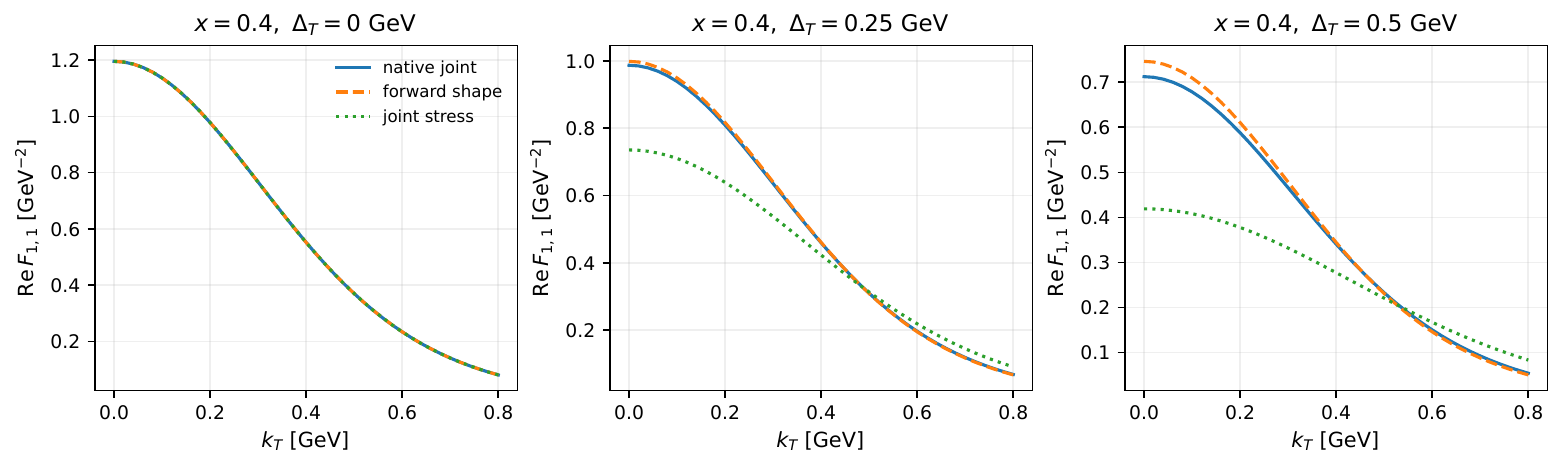}
\caption{Joint-model sensitivity of $F_{1,1}$ at $x=0.4$, $\bm k\perp\bm\Delta$, as functions of $|\bm k|$ at three fixed transfers: native joint, forward shape, and the illustrative nonforward stress of Eq.~\eqref{eq:blind}. State, current, width, matching tables, and fitted kernels are common. The curves agree forward and separate nonforward; the separation is the joint freedom the present inputs leave open.}
\label{fig:f11-model-surfaces}
\end{figure*}

\begin{figure*}[t]
\centering
\includegraphics[width=0.95\textwidth]{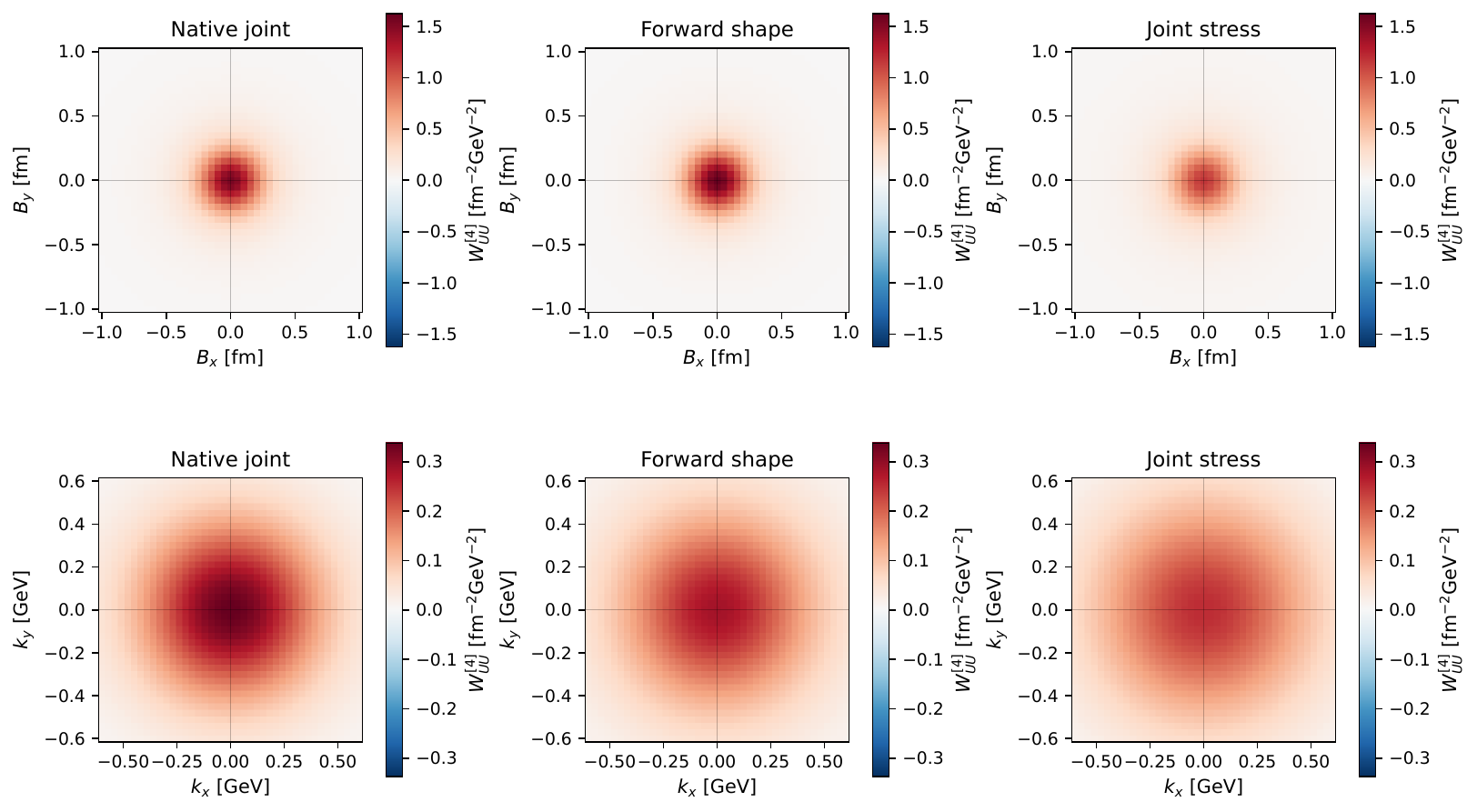}
\caption{The same three joint prescriptions compared through finite-window tomography at $x=0.4$. Top: impact planes at $\bm k=(0.4,0)\GeV$. Bottom: momentum planes at $\bm B=(0.5,0)\,{\rm fm}$. All use $|\bm\Delta|\leq4\GeV$ and the same state and kernels. Each row has a common color scale. The differences between the maps show model dependence only.}
\label{fig:wigner-model}
\end{figure*}

Figures~\ref{fig:f11-model-surfaces} and \ref{fig:wigner-model} show what the open joint direction does in phase space. All three prescriptions satisfy the same input criteria and give different off-forward and impact-space structures. The geometric representation is faithful to each of them. Choosing among them is a job for a nonforward measurement, and these figures show what such a measurement would be looking at.

The dense evaluations pass 476 numerical checks. The largest radial-quadrature and evolution-grid discrepancies are $4.19\times10^{-9}$ and $1.26\times10^{-10}$ in the relevant displayed units, compared with separate absolute tolerances of $2\times10^{-7}$ for $F$ in $\GeV^{-2}$ and $W^{\rm fm}$ in $\mathrm{fm}^{-2}\GeV^{-2}$. Transfer-quadrature refinement is below $5.4\times10^{-15}$, and the tested intrinsic-separation tail change is below $4.5\times10^{-16}$ in those readouts. An independent direct transfer-angle calculation agrees at 100 selected map/window/model values to $1.8\times10^{-15}$. The finite readouts are therefore numerically converged to the stated precision.

\subsection{Mixed position--momentum projections}
To display the third type of unpolarized panel in Ref.~\cite{chakrabarti}, we define an explicit finite-$k_x$ projection,
\begin{align}
 \mathcal Q_{K,D_{\max}}(x,B_x,k_y)
 ={}&\int_{-D_{\max}}^{D_{\max}}\frac{\dd\Delta_x}{2\pi}
 e^{-\ii\Delta_x B_x}
 \nonumber\\
 &\times\int_{-K}^{K}\dd k_x\,
 \F(x,k_x,k_y,\Delta_x,0).
 \label{eq:mixed}
\end{align}
For a hard circular transfer window, this equals the full-$B_y$ projection of $W_w$ in the Fourier-projection sense, followed by the finite momentum acceptance. The reference uses $x=0.4$, $K=0.8\GeV$, and $D_{\max}=4\GeV$. The displayed density is $\mathcal Q^{\rm fm}=\mathcal Q/(\hbar c)$ with units $\GeV^{-1}{\rm fm}^{-1}$, because only one position coordinate remains.

\begin{figure*}[t]
\centering
\includegraphics[width=\textwidth]{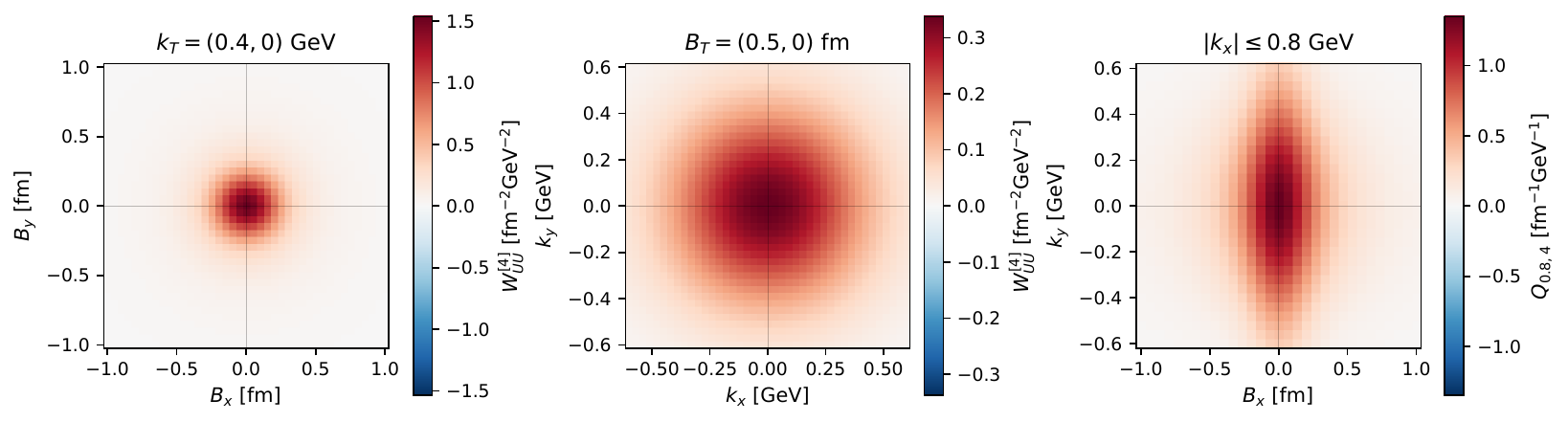}
\caption{Three complementary fixed-$x=0.4$ unpolarized readouts, in a layout analogous to the $UU$ projections of Ref.~\cite{chakrabarti}. Left: impact plane at $\bm k=(0.4,0)\GeV$. Center: momentum plane at $\bm B=(0.5,0)\,{\rm fm}$. These two display $W_{w_4}/(\hbar c)^2$. Right: mixed $(B_x,k_y)$ projection of Eq.~\eqref{eq:mixed}, with $K=0.8\GeV$ and $D_{\max}=4\GeV$, displaying $\mathcal Q/(\hbar c)$. Its color units therefore differ from the first two panels. The finite windows are part of the definitions.}
\label{fig:three-projections}
\end{figure*}

Figure~\ref{fig:three-projections} puts the three readouts alongside one another without equating their normalizations. The mixed panel uses a $31\times31$ grid. The full calculation includes $K=1.2$ and $1.6\GeV$ alternatives, extending its tested integration domain to $|\bm k|\leq\sqrt{1.6^2+0.6^2}=1.709\GeV$. Radial, evolution-grid, transfer-order, $k_x$-order, and intrinsic-tail comparisons pass 152 checks. Their largest discrepancies are $2.07\times10^{-9}$, $3.01\times10^{-11}$, $2.89\times10^{-15}$, $1.17\times10^{-12}$, and $2.23\times10^{-16}\GeV^{-1}{\rm fm}^{-1}$, respectively.

An independent calculation performs the finite $k_x$ integral analytically in intrinsic-separation space, replacing its exponential by $2K\,{\rm sinc}(Kr_x/\pi)$, then integrates the full azimuth and both signs of transfer directly. It agrees with 100 values of the main calculation across the model and window alternatives to $3.47\times10^{-13}\GeV^{-1}{\rm fm}^{-1}$. Here $\operatorname{sinc}(z)=\sin(\pi z)/(\pi z)$. This check bypasses both the $k_x$ quadrature of the main calculation and its Hankel implementation.

\begin{figure*}[t]
\centering
\includegraphics[width=\textwidth]{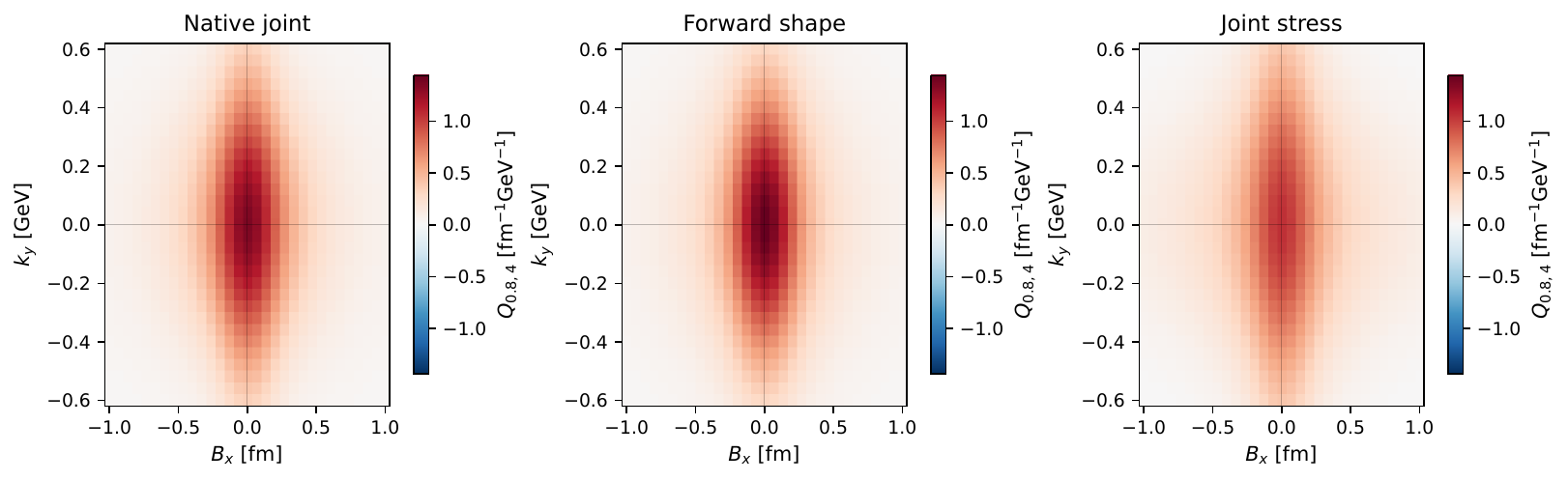}
\caption{Mixed projection at $x=0.4$, $|k_x|\leq0.8\GeV$, and $|\Delta_x|\leq4\GeV$ for the three joint-model prescriptions. All panels use the same state, current, kernel table, and matching inputs, and share one color range. The density is $\mathcal Q/(\hbar c)$. The differences show the model dependence.}
\label{fig:mixed-model}
\end{figure*}

\begin{figure*}[t]
\centering
\includegraphics[width=0.85\textwidth]{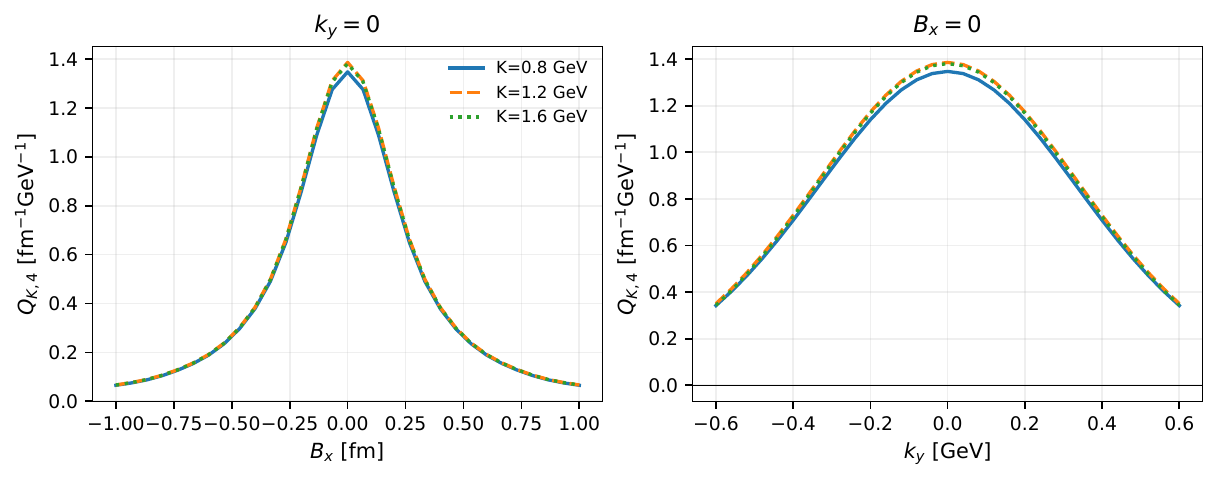}
\caption{Dependence of the mixed projection on its finite momentum acceptance, $K=0.8,1.2,1.6\GeV$, at $x=0.4$ and $D_{\max}=4\GeV$. The cuts shown use the native-joint closure.}
\label{fig:mixed-window}
\end{figure*}

Figures~\ref{fig:mixed-model} and \ref{fig:mixed-window} separate joint-model dependence from finite-acceptance dependence. Neither is absorbed into the conditional kernel profile of Sec.~\ref{sec:profiles}. The mixed construction is a useful additional view; the all-momentum limit, which requires ultraviolet renormalization, stays outside the finite readout.

For scale, the primary mixed-map peak is $1.347\GeV^{-1}{\rm fm}^{-1}$. Raising $K$ from $0.8$ to $1.2$ or $1.6\GeV$ changes the map by at most $0.039$ or $0.034\GeV^{-1}{\rm fm}^{-1}$, respectively, over the displayed grid. These correspond to $2.9\%$ and $2.5\%$ of the primary peak. They need not grow monotonically with acceptance, since the integrand is a signed correlator. The finite-acceptance dependence is small next to the joint-model dependence, which is once again the dominant freedom.

\section{From the proton to neutron, deuteron, and \texorpdfstring{spin-$3/2$}{spin-3/2} targets}
\label{sec:extensions}
The proton construction was built to be composed. The shared state and its registers, the operator definitions, the symmetry-resolved network with its exact color, charge, and exchange structure, the geometric feature machinery, the matching conventions, and the treatment of inputs all carry over to the next targets. What each target adds is stated explicitly: new amplitudes, new currents, and a higher polarization rank, each with its own validation. The tensor-network organization turns this into a sequence of controlled steps instead of a new model for every target. Composing two subunits in a symmetry-resolved network is a tensor product of their charge-labeled registers with a fixed intertwiner. The exchange, parity, and charge constraints of the composite are inherited rather than reimposed, and the one- and two-body operator hierarchy acts on the registers the proton calibration has already validated. In a hand-built parametrization each of these constraints would be a separate condition to enforce, and their number grows multiplicatively with the spin--flavor--color register.

\subsection{Neutron: isospin transport with independent checks}
The first step is a joint proton--neutron description. In the charge-symmetric limit, an isospin rotation exchanges the $u$ and $d$ responses,
\begin{equation}
 F^{u/n}=F^{d/p},\qquad F^{d/n}=F^{u/p},
 \label{eq:isospin}
\end{equation}
for corresponding flavor operators and common conventions. In the shared construction the relation is implemented with isoscalar and isovector operator blocks acting on the same state, with electromagnetic charges applied only at the observable layer. Because the network bonds carry $I_z$ labels, the transport is a relabeling, not a refit. Equation~\eqref{eq:isospin} is a charge-symmetry approximation, and quark-mass and electromagnetic corrections enter as additional terms when required.

The tests are charge normalization, proton--neutron flavor exchange, elastic and axial responses, and consistency of any nuclear corrections used to obtain neutron information. Inputs that were themselves derived from a proton--neutron model are marked as such. The neutron stage tells us whether a common nucleon construction transfers across isospin before nuclear composition adds further freedom. If it fails, the failure points to a missing isovector current instead of disappearing into refitted parameters.

\subsection{Deuteron: composition and tensor polarization}
The spin-$1$ deuteron adds relative nucleon motion, two-body currents, and tensor polarization, and its GPD structure is already qualitatively richer than that of a nucleon~\cite{berger}. The nuclear construction is
\begin{align}
 |d,\Lambda\rangle={}&\sum_{\lambda_p\lambda_n}
 \int[\dd\Gamma]\,
 \Psi^{\Lambda}_{\lambda_p\lambda_n}(\Gamma)
 |p,\lambda_p;n,\lambda_n;\Gamma\rangle,
 \nonumber\\
 \mathcal O_d={}&\mathcal O_p\otimes I+I\otimes\mathcal O_n
 +\mathcal O_{2N}.
 \label{eq:deuteron}
\end{align}
Here $[\dd\Gamma]$ includes a specified relative-momentum measure and the appropriate boosts. Equation~\eqref{eq:deuteron} is the operation the symmetry-resolved network was designed for. The proton and neutron subunits enter as validated registers, the deuteron's $J=1$ and $I=0$ labels fix the intertwiner, the one-body terms carry the calibrated nucleon structure, and the two-body term is a separately labeled operator that can be switched on and tested. It is kept whenever an identity or an observable requires it.

For continuity with the broader construction, a chosen AV18 relative wave function~\cite{av18} is held fixed during the first transfer test, so that changes in the embedded nucleon response are distinguished from changes in nuclear dynamics. Boost and spin-rotation prescriptions, off-shell continuation, exchange currents, and non-nucleonic components are separate assumptions, each stated explicitly. The tests are spin-$1$ charge and magnetic normalization, quadrupole observables, polarization projectors, and the relevant rotational conditions within the approximation used; an impulse-only result is labeled as such. Since the spin density matrix enters the shared matrix elements linearly (Sec.~\ref{sec:shared-carrier}), tensor polarization is a projection of the same composite construction and not a new model.

\subsection{\texorpdfstring{Spin-$3/2$}{Spin-3/2}: new multipoles and operator bases}
The final stage is a general spin-$3/2$ target. The most interesting applications are nuclei, $^{7}$Li in particular, for which polarized targets exist and whose many-body dynamics and currents come from the same composition machinery. A $\Delta$ baryon would in addition require its resonance character to be treated explicitly.

The hierarchy is transparent in the spin-density decomposition
\begin{equation}
 \rho^{(J)}=\sum_{K=0}^{2J}\sum_{Q=-K}^{K}
 t_{KQ}\,T^{(K)}_Q.
 \label{eq:multipoles}
\end{equation}
Spin $1/2$ allows scalar and vector polarization, spin $1$ adds rank-two tensor polarization, and spin $3/2$ adds rank-three octupole polarization. The dimensions $1+3+5+7=16$ describe the spin-$3/2$ density-operator space; the number of independent GTMD scalar functions follows from the full covariant bilocal parametrization, discrete-symmetry analysis, and independent helicity projectors. Existing spin-$3/2$ GPD and transversity classifications provide boundary checks~\cite{fu2022,fu2024}, and the complete leading-twist spin-$3/2$ TMD operator basis, with its Born response maps, is available to serve as the observable layer.

On the architectural side, the larger target-spin registers and symmetry intertwiners are introduced only after the operator basis is fixed, so that rank requirements can be measured on the actual response matrices with the sizing rule of Sec.~\ref{sec:kernel-network}. Quantum feature circuits, classical equivariant nodes, and hybrid alternatives are then compared at matched output accuracy and full execution cost, including state preparation, contractions, readout, and derivatives. The response matrices of a rank-three target are large enough for this comparison to be meaningful.

\begin{table}[t]
\caption{Staged program. The proton construction and readouts of this paper are the validated first step; each later stage reuses the components in the second column and adds the requirements in the third.}
\label{tab:roadmap}
\begin{ruledtabular}
\begin{tabular}{p{0.16\columnwidth}p{0.34\columnwidth}p{0.39\columnwidth}}
Target & Reusable component & New requirement \\
\hline
Proton & Shared carrier, currents, geometric nodes & Matched $F_{1,1}$ with explicit closure; this paper \\
Neutron & Nucleon operators and isospin blocks & Charge-symmetry tests and neutron-specific inputs \\
Deuteron & Validated proton/neutron subunits & Relative motion, two-body currents, rank-two polarization \\
Spin $3/2$ & Symmetry and learning machinery & New helicity basis, rank-three polarization, target-specific dynamics
\end{tabular}
\end{ruledtabular}
\end{table}

\section{Discussion}
\label{sec:discussion}
We have built a calibrated proton GTMD that takes in experimental, lattice, and phenomenological information through common physical degrees of freedom, and we have said quantitatively what that information determines. Operator and matching conventions fix the meaning of the selected $F_{1,1}$. Shared amplitudes and currents connect the source projections. Normalized transverse kernels and a declared joint closure complete the model, and quantum and classical geometric nodes give tested computational representations of the resulting functions. At every step the route from an input to the constructed correlator is explicit.

The physics comes down to two statements. Forward transverse information, entered with its full covariance, narrows the attainable GTMD by about $81\%$ within the kernel family used; this is a direct measure of how much a modern TMD extraction tells us about the five-dimensional correlator. The joint dependence on quark momentum and momentum transfer is a different matter. A deformation invisible to every present input changes the nonforward value by $36\%$, which is more than the whole conditional range the inputs allow. Both statements are conditional on the stated family and closure. The deformation is given explicitly, and its effect on the momentum-space and phase-space images is shown in Sec.~\ref{sec:tomography}.

The second statement sets the agenda. A useful future constraint has to respond to a deformation like Eq.~\eqref{eq:blind}, through a validated process map or an explicit model relation. The candidates are matched lattice bilocals and hard exclusive channels whose factorization is established for the specific operator and kinematics. In the shared-state construction such a constraint can be added without disturbing the forward calibration, because it enters as one more projection of the same state. Nonzero skewness, polarized channels, and orbital-angular-momentum observables are further projections of the same construction, each with its own operator and matching work.

The present results also identify two model-development steps. The large-transfer pole found in the aggregate closure goes away in the modewise closure, which restores a single state-independent effective operator on the finite carrier; adopting it requires renewed source fits and is the first task of the next iteration. The inherited evolution ratio $R_{\rm src}$ can be replaced by a native momentum-space evolution operator at a single perturbative order, removing the mixed perturbative order of the present construction.

Finally, the architecture was built for the targets that follow. In the neutron, the deuteron, and spin-$3/2$ nuclei, the properties that the proton construction has established, exact color contraction, conserved charges on every bond, exchange statistics verified on the reconstructed maps, and a clean separation between reorganizing, truncating, and enlarging a representation, become the mechanism of composition itself. Isospin transport is an operator relabeling; nuclear binding is a tensor product of validated subunits with a labeled one- and two-body operator hierarchy; higher polarization ranks are larger intertwiners whose rank requirements are measured on the actual response matrices. Going proton-first lets each physical question be settled before the next layer of complexity is added. The proton construction presented here is the first step.

\section*{Data and code availability}
The analysis code and the numerical data behind the figures are available from the author upon reasonable request.

\section*{Use of AI tools}
Overleaf's built-in AI tools, which mainly use OpenAI models together with Writefull's proprietary language models, were used for proofreading and editing. Claude code (Anthropic) was used for editing and refining the script that generates Fig.~\ref{fig:geometric-nodes}. The Codex coding agent (OpenAI) was also used to organize the analysis workflow, to maintain content-hash records of its inputs and outputs, and to run validation checks on the analysis code (Appendix~\ref{app:support}). The author has checked all content and takes full responsibility for it.

\appendix
\section{Calibration procedure and treatment of inputs}
\label{sec:inference}
\subsection{Inputs and compatibility criteria}
This appendix describes the calibration procedure behind the reference fit, so that the numbers in the main text can be traced. Its numerical criteria diagnose how the common model responds to the selected information; because several inputs rest on common data and their joint covariance is incomplete, they are compatibility diagnostics rather than a global likelihood. The physical organization of the inputs is described in Sec.~\ref{sec:information}.

The construction uses one shared state and shared current responses. Each input enters with its published central values, its covariance on the supported subspace, and its own nuisance parameters. We do not rescale errors to make the final objective appear acceptable. If $d_a$ is a source vector, $f_a(\theta)$ its response, $V_a$ its supported covariance, and $A_a\lambda$ a permitted shared shift, its quadratic contribution has the schematic form
\begin{equation}
 Q_a=\bigl[f_a(\theta)-d_a-A_a\lambda\bigr]^T
 V_a^{+}\bigl[f_a(\theta)-d_a-A_a\lambda\bigr],
 \label{eq:source}
\end{equation}
with each nuisance penalty counted once. Here $V_a^+$ denotes the inverse on the supported subspace. Source-specific logarithmic terms and multiplicative-systematic conventions are retained when required. They are not replaced by independent point errors.

The ART block consists of $110$ correlated projections of the extracted distributions, derived from $642$ replicas of that extraction. It constrains both $u$- and $d$-valence responses even though the physical target displayed here is $u$ valence. These are projections of a phenomenological extraction~\cite{art25}, not $110$ new independent measurements. A selection of $39$ HERA points enters through the conditional DIS map validated previously. The remaining $54$ inputs comprise three derived Mainz axial nodes~\cite{mainz}, 17 ETMC vector and nine ETMC axial-second-moment rows~\cite{etmc}, four flavor-diagonal PNDME charges~\cite{pndme}, seven Christy cross sections~\cite{christy}, four Kubon neutron magnetic-form-factor extractions~\cite{kubon}, eight Zhan proton polarization ratios~\cite{zhan}, and two static magnetic moments. The row count overstates the number of independent degrees of freedom, which is one reason the criteria are read as diagnostics.

The Mainz nodes retain the covariance inherited from their common source parametrization. The ETMC and PNDME joint covariances are not available in the published results we use; diagonal marginal-error blocks and zero cross-study correlations are explicitly conditional assumptions, not published independence statements. ETMC vector and axial rows share an ensemble. Zhan's shared-systematic scenario is likewise conditional rather than a recovered experimental covariance. Christy's $1.7\%$ normalization uncertainty is assigned a single penalized nuisance parameter, and Kubon's reported total already contains its extraction-theory contribution. The effective axial, tensor, and moment maps retain their earlier conditional matching assumptions; validating the vector GTMD does not complete their QCD matching. These conditional assumptions are recorded so that they can be replaced as the corresponding covariances become available.

HERA data already enter the collinear distributions used by the TMD extraction. Without the cross-covariance or an explicit analysis of the common data, multiplying the two source likelihoods would double count information. We instead minimize the ART compatibility criterion subject to caps on the other criteria,
\begin{align}
 &\underset{\theta,p}{\operatorname{minimize}}\ Q_{\rm ART}(\theta,p),
 \nonumber\\
 &Q_{\rm HERA}(\theta)\leq203.07,
 \nonumber\\
 &Q_{54}(\theta)\leq147.96,
 \quad p_m\in\Delta_{207},
 \label{eq:fit}
\end{align}
alongside normalization, operator, and domain checks. These caps are taken from an earlier stage of the construction; they are practical tolerances rather than likelihood-ratio thresholds. Criteria and caps are quoted to the precision needed for the comparisons. The residuals of the individual inputs, including those that worsen within the aggregate cap, are reported in Appendix~\ref{sec:results}.

\subsection{Alternating state and kernel steps}
At fixed shared state, current coefficients, and width, all ART predictions are affine in the $4\times208$ kernel weights. The kernel step is therefore a convex quadratic problem on a product of simplices. The solutions we keep are normalized exactly using rational weights and re-evaluated with high-precision arithmetic. An affine supporting bound quantifies the finite-table optimization gap. It cannot bound errors in the underlying evolution tables or prove optimality over a continuous Gaussian-width family.

The state step varies $113$ real tangent coordinates of the normalized native $114$-component state, six source/current coordinates, and the common width. Its dependence is nonlinear, and the aggregate physical response includes the ratio in Eq.~\eqref{eq:native}. The state we report is feasible and passes all checks, but neither constrained stationarity nor global optimality is established. We alternate state updates and kernel polishing and keep the best solution that passes the checks; termination of the solver is not by itself taken as an acceptance criterion.

The fitting loop sits inside a sequence of decisions about the architecture. We first ask whether the current shared response can represent the physics included, and examine the residuals of each input and the directions to which the inputs are insensitive. We then ask whether a missing symmetry coupling, physical tensor factor, or response operator calls for a change of the network, testing exact reorganizations separately from bond truncation and from an enlarged physical family. Finally we check whether the classical, quantum, or hybrid node lacks the feature rank or numerical conditioning needed to represent the resulting response. After each change the physical and input checks are repeated. The steps can be interleaved when a limitation of the representation is already evident, but the same checks are required.

A new hidden node is justified by a demonstrated representation or optimization need; a poor fit may instead reflect inconsistent input assumptions, a deficient physical response, or a constraint the data do not provide, and the checks above are meant to distinguish these cases. Exact topology reorganization can change contraction cost or conditioning without enlarging the physical family. The matched calculation reported here required no further topology change: its improvement came from matching, shared-state/current/width variation, and normalized kernels within the already validated representation, as the rank analysis of Sec.~\ref{sec:kernel-network} anticipated.

\subsection{Conditional profiles and identifiability}
\label{sec:profiles}
For a fixed final state, let $a_z^Tp$ be a target readout at kinematics $z$. The admissible kernel set is
\begin{equation}
 \mathcal P_\epsilon=\left\{p_m\in\Delta_{207}:
 Q_{\rm ART}(p)\leq(1+\epsilon)Q_{\rm ART}^{\rm ref}\right\}.
 \label{eq:profileset}
\end{equation}
We minimize and maximize $a_z^Tp$ over this set. The outer allowance is $\epsilon=0.01$. To leave a numerical margin, the reported inner endpoints use $\epsilon=0.008$. The reference criterion is $5.056\times10^{5}$, giving inner and outer caps $5.096\times10^{5}$ and $5.106\times10^{5}$. These allowances are not confidence levels.

For a single target, convex interpolation between feasible endpoint kernels remains feasible in the same fixed numerical table. Endpoints for different targets usually require different kernels, so their collection is not a simultaneously attained curve. The set also holds the state, width, current, longitudinal basis, matching prescription, and closure fixed. Its width is a conditional sensitivity within the stated family.

The distinction between projection and joint information can be expressed locally through the source Jacobian $J_{a\alpha}=\partial f_a/\partial\theta_\alpha$. Directions $v$ satisfying $Jv=0$ are locally unidentified by those sources. Equation~\eqref{eq:blind} gives a stronger finite example: the forward responses used in the fit remain exactly invariant while the nonforward target changes. Restricting such a direction by a closure narrows the predictions, but the restriction is model input rather than data, and we keep the two apart.

\section{Calibration diagnostics and conditional sensitivities}
\label{sec:results}
\subsection{Compatibility improvement and residual tension}
This appendix collects the numerical details of the fit. The values of the criteria refer to the calibrated model used throughout the paper.

\begin{figure*}[t]
\centering
\includegraphics[width=\textwidth]{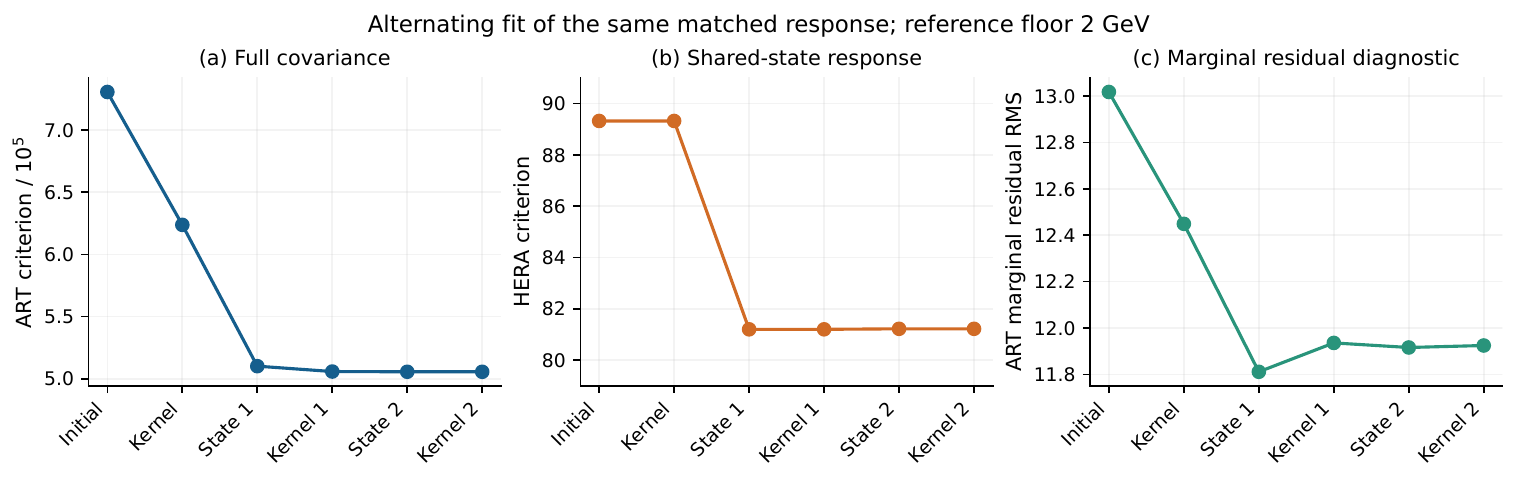}
\caption{Alternating shared-state and kernel fit under the reference physical response. (a) Full-covariance ART criterion. (b) HERA criterion, which changes only in the state steps. (c) ART marginal residual RMS computed with diagonal errors. The last kernel polish decreases (a) while slightly increasing (c), illustrating the effect of correlations. The criterion $Q_{54}$ for the remaining 54 inputs stays within its cap throughout. These criteria are not summed into an independent joint likelihood.}
\label{fig:fit-sequence}
\end{figure*}

Figure~\ref{fig:fit-sequence} shows the alternating sequence of the final fit within the same matched response. The initial shared state and initial kernels give $Q_{\rm ART}=7.31\times10^{5}$. Kernel fitting lowers this to $6.24\times10^{5}$; two state updates and their kernel polishes reach $5.06\times10^{5}$. The improvement is approximately $31\%$ relative to the initial response. The last kernel polish improves the objective by only $0.0047\%$, much less than the changes induced by transition-model alternatives below. No further state cycle is included.

The final criteria for HERA and for the remaining 54 inputs are $81.22$ and $147.96$, respectively. Preserving the latter total does not imply that every source improves: the PNDME spin block increases by $2.61$ relative to the initial state, compensated by improvements elsewhere.

\begin{figure}[t]
\centering
\includegraphics[width=\columnwidth]{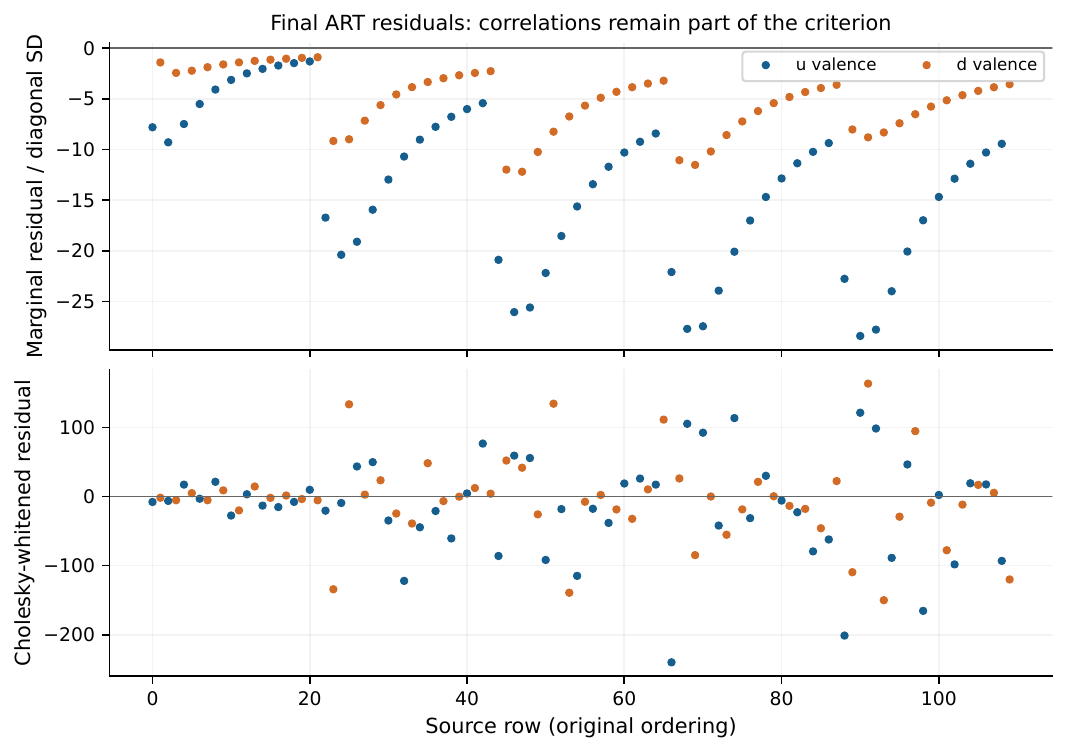}
\caption{Final ART residuals in the original source ordering. Top: marginal residuals divided by diagonal standard deviations. Bottom: Cholesky-whitened residuals using the full supported covariance. The $u$ and $d$ labels denote the original rows; after whitening, each component can mix correlated source information. Both panels show that substantial tension remains.}
\label{fig:art-residuals}
\end{figure}

The remaining ART discrepancy is substantial. The diagonal-error residual RMS is $11.92$, and the largest absolute marginal residual is $28.40$. The full-covariance criterion cannot be interpreted from that marginal RMS alone. Figure~\ref{fig:art-residuals} displays both marginal and Cholesky-whitened residuals; the whitened components depend on the retained ordering, while their squared norm gives the quadratic criterion. A lower objective establishes improvement under the stated response and covariance; the remaining tension is the diagnostic that points to the matching and closure refinements identified in Sec.~\ref{sec:discussion}.

\begin{figure*}[t]
\centering
\includegraphics[width=0.86\textwidth]{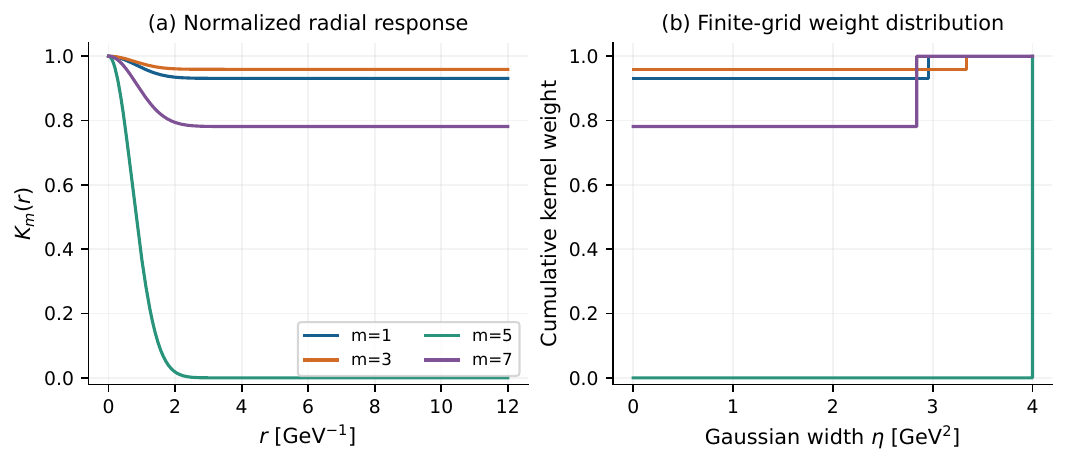}
\caption{Final four normalized transverse kernels. (a) Intrinsic-separation dependence. (b) Cumulative weights on the fixed 208-point Gaussian-width grid. The labels $m=1,3,5,7$ are longitudinal modes. Positivity and finite support are model restrictions on the kernels and do not assert positivity of the GTMD. The separation $r$ is conjugate to $k$ and is distinct from impact parameter $B$.}
\label{fig:kernels}
\end{figure*}

The normalized kernels in Fig.~\ref{fig:kernels} are shared across all source rows assigned to a mode. Their cumulative weight distributions display the fitted finite support without treating a set of isolated Gaussian widths as measured partonic components. The final hybrid node reproduces these four probability rows after the interior replacement of exact zeros described in Sec.~\ref{sec:kernel-network}. This is the check that the node reproduces the fitted response functions.

\subsection{What has been constrained}
The original validation covers $x=0.2,0.4,0.6$, $\bm k=(0,0),(0.2,0),(0,0.2),(0.4,0)\GeV$, and $\bm\Delta=(0,0),(0.2,0),(0.5,0)\GeV$: 36 targets for each aggregate closure. All 144 minimum and maximum endpoints were recomputed independently. The largest attained-inner-endpoint to supporting-bound gap is $3.32\times10^{-5}\GeV^{-2}$. Results obtained with the alternative evolution grid satisfy the outer allowance, not necessarily the inner one.

\begin{table*}[t]
\caption{Physical-model anchors at $x=0.4$, $|\bm k|=0.2\GeV$. Nonforward entries use $\bm k\perp\bm\Delta$, $|\bm\Delta|=0.5\GeV$. Values are in $\GeV^{-2}$. Ranges vary finite kernels at fixed state and prescription.}
\label{tab:anchors}
\begin{ruledtabular}
\begin{tabular}{lccc}
Kinematics & Central & Inner range & Before ART cap\\
\hline
Forward & $0.978$ & $[0.813,0.997]$ & $[0.060,1.033]$\\
Joint & $0.588$ & $[0.499,0.599]$ & $[0.037,0.620]$\\
Forward shape & $0.610$ & $[0.507,0.622]$ & $[0.037,0.645]$
\end{tabular}
\end{ruledtabular}
\end{table*}

\begin{figure*}[t]
\centering
\includegraphics[width=\textwidth]{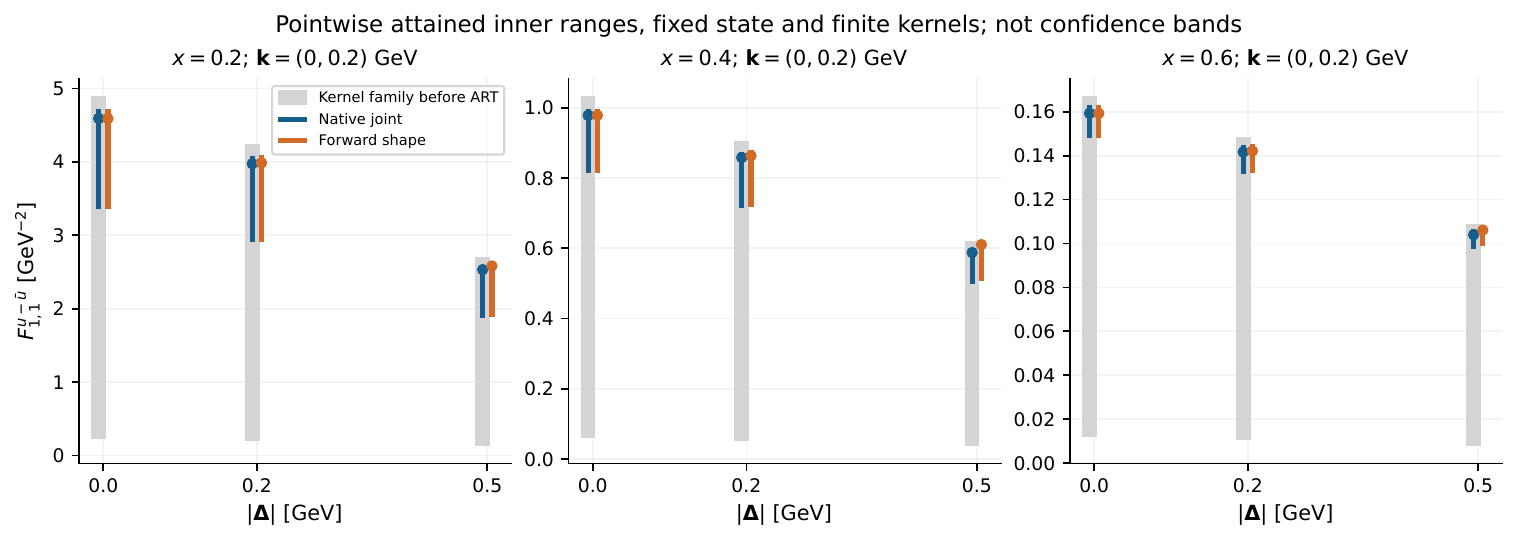}
\caption{Conditional kernel ranges at three longitudinal fractions, with $\bm k=(0,0.2)\GeV$ and $\bm\Delta$ along the horizontal momentum axis. Gray segments show the native-joint finite-kernel range before imposing the ART cap. Colored segments are attained inner ranges for the two closures; dots are their central values. Small horizontal offsets separate coincident transfer coordinates. Each endpoint is a separate optimized kernel table, so the segments do not form a single jointly attained uncertainty band.}
\label{fig:profiles}
\end{figure*}

Table~\ref{tab:anchors} gives the representative anchors. The attained width decreases by $81\%$ forward and $83\%$ nonforward relative to the same finite-kernel family without the ART cap. Figure~\ref{fig:profiles} shows the corresponding transfer-dependent restrictions for three $x$ values. The narrowing is a direct consequence of the source criterion within the stated family.

The ranges should not be read as global. For an earlier fitted state the forward kernel range was $[0.718,0.940]$, whereas the final central forward value is $0.978$. Changing the shared state moves the result outside that earlier conditional range; each state defines its own range.

The joint stress of Eq.~\eqref{eq:blind} gives a second example. At the nonforward anchor it changes $0.588$ to $0.378$, retaining the forward value and all current input criteria. The narrower profile therefore coexists with a large, explicitly exhibited joint freedom: small-$r$ matching fixes the leading operator connection, the data constrain the forward response directions, and the closure supplies the rest.

\subsection{Separate matching and model sensitivities}
\begin{figure*}[t]
\centering
\includegraphics[width=0.93\textwidth]{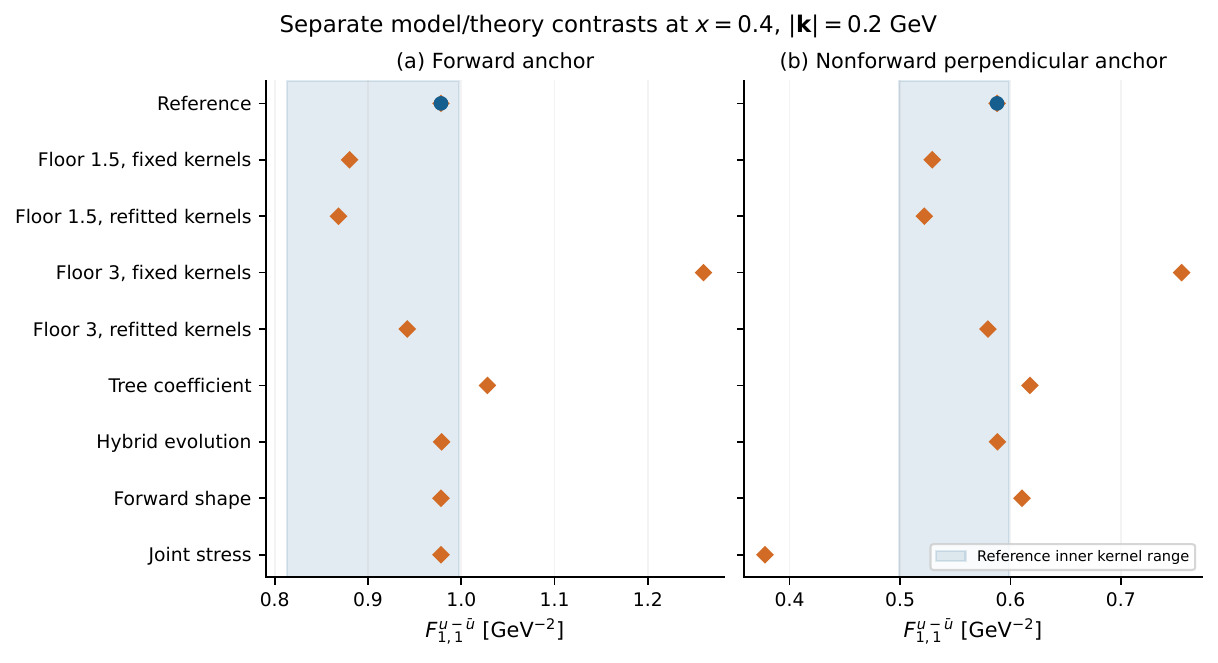}
\caption{Separate theory and model contrasts at the two $x=0.4$ anchors. Shading is the reference inner finite-kernel range, with no statistical coverage interpretation. Diamonds show fixed-kernel changes except where explicitly labeled as alternative-floor kernel refits. The joint stress leaves forward responses invariant. Tree coefficient means removal of the one-loop matching coefficient while retaining the evolved boundary; it is not a uniformly LO calculation. The contrasts are not independent random uncertainties and are not combined in quadrature.}
\label{fig:model-sensitivity}
\end{figure*}

Figure~\ref{fig:model-sensitivity} separates changes in the transition floor, coefficient order, evolution prescription, and joint closure. At fixed final kernels, floors $1.5$, $2$, and $3\GeV$ give forward values $0.880$, $0.978$, and $1.260$ and nonforward native-joint values $0.529$, $0.588$, and $0.755$. A separate kernel refit at floor $3\GeV$ moves the latter to $0.580$, showing that the intrinsic shape can absorb much of this particular change. At floor $1.5\GeV$, the refitted value is $0.522$.

The floor-$1.5\GeV$ criterion is lower than the reference even before refitting: $3.39\times10^{5}$, compared with $5.06\times10^{5}$; after kernel refitting it is $3.27\times10^{5}$. The reference floor was fixed in advance and is not the optimum over transition prescriptions. A lower source criterion does not by itself establish the physical accuracy of a regulator choice. These shifts are much larger than the last kernel-polish improvement.

The tree-coefficient contrast changes the forward/nonforward anchors to $1.028/0.618$. The hybrid-evolution contrast, using one-loop evolution anchored to the source rapidity kernel, gives $0.979/0.588$. The modewise model of Eq.~\eqref{eq:modewise} was tested at the original and first-cycle states, giving ART criteria $6.25\times10^{5}$ and $5.07\times10^{5}$, respectively. It does not improve the corresponding aggregate fits. At those identical states and kernels, the maximum change among the 36 targets is approximately $0.0125\GeV^{-2}$. These tests were not repeated at the final second-cycle state.

\subsection{Numerical validation versus physical accuracy}
The 36-point Fourier maps of the final fit satisfy absolute tolerances of $2\times10^{-7}\GeV^{-2}$ for the radial quadrature, the evolution grid, and the tested tail. Radial discrepancies are below $4.8\times10^{-8}$ and evolution-grid differences below $1.2\times10^{-9}\GeV^{-2}$. Finite angular harmonics were compared with independent angular quadrature. These are tested numerical convergence statements for the finite construction.

The finite-table kernel optimization gap is approximately $2.37\times10^{-5}$ in the ART criterion, whereas the change from the alternative evolution grid is approximately $0.0102$. Numerical precision is therefore far below the model-assumption and source sensitivities, which are the quantities that govern the physical accuracy of the construction. The dense readouts in Sec.~\ref{sec:tomography} have separate domain and integration checks.

\section{Endpoint prescription and operator conventions}
\label{app:plus}
For a smooth test function $h$, the denominator plus distribution obeys
\begin{equation}
 \int_0^1\dd y\,\left[\frac1{1-y}\right]_+h(y)
 =\int_0^1\dd y\,\frac{h(y)-h(1)}{1-y}.
\end{equation}
Thus the logarithmic convolution in Eq.~\eqref{eq:coefficient} is implemented as
\begin{align}
 &\int_x^1\frac{\dd y}{y}(1+y^2)
 \left[\frac1{1-y}\right]_+ H(x/y)
 \nonumber\\
 &=\int_x^1\dd y\,
 \frac{(1+y^2)H(x/y)/y-2H(x)}{1-y}
 \nonumber\\
 &\quad+2H(x)\ln(1-x).
 \label{eq:plus-implementation}
\end{align}
This differs from placing a plus sign around the complete ratio $(1+y^2)/(1-y)$; changing that notation without the compensating delta term would change the coefficient. Constant and polynomial test functions, endpoint limits, and independent quadrature were included in the tests of the coefficient.

The transverse spin-tensor convention uses explicit Euclidean components $k_i\sigma^{i+}$ and $\Delta_i\sigma^{i+}$. A literal covariant contraction $\sigma^{k_Tn}=k_{T\mu}n_\nu\sigma^{\mu\nu}$ carries an additional minus sign for each single transverse contraction in our metric. Companion coefficients $F_{1,2}$ and $F_{1,3}$ must be converted with that sign if changing basis. $F_{1,1}$ and the half-trace result are unchanged. Preserving this distinction prevents a correct scalar trace from concealing an inconsistent helicity matrix.

The formal Fourier inversion in Eq.~\eqref{eq:fourier} is used at finite $k$; the all-$k$ integral relating the renormalized GTMD to a collinear GPD requires ultraviolet matching and distributional subtractions, which the finite numerical transforms of the effective closure do not attempt.

\section{Fourier harmonics, units, and finite transfer windows}
\label{app:transforms}
Write the intrinsic-separation dependence at a fixed transfer direction as
\begin{equation}
 \Ft(x,r,\phi_r;D)=\sum_{n=-4}^{4}s_n(x,r,D)e^{\ii n\phi_r},
 \qquad D=|\bm\Delta|.
\end{equation}
The finite harmonic content follows from the retained degree-four transverse polynomial in the effective overlap, not from a general theorem for a QCD GTMD. The first transform is
\begin{equation}
 F(x,k,\phi_k;D)=
 \sum_{n=-4}^{4}\ii^n e^{\ii n\phi_k}
 \int_0^\infty\frac{r\,\dd r}{2\pi}J_n(kr)s_n(x,r,D).
 \label{eq:first-hankel}
\end{equation}
The angular identity is checked against direct complex quadrature. Global rotations of $\bm k$ and $\bm\Delta$, transfer-reversal Hermiticity, and scalar reality are tested separately.

For a Fourier expansion in transfer angle,
\begin{equation}
 F(x,\bm k,D,\phi_\Delta)=\sum_n f_n(x,\bm k,D)e^{\ii n\phi_\Delta},
\end{equation}
the finite-window impact transform is
\begin{align}
 W_w
 &=
 \int\frac{\dd^2\bm\Delta}{(2\pi)^2}
 e^{-\ii\bm\Delta\cdot\bm B}
 w(D) F(x,\bm k,\bm\Delta)
 \nonumber\\
 &=\sum_n(-\ii)^n e^{\ii n\phi_B}
 \nonumber\\[-2pt]
 &\quad\times
 \int_0^{D_{\max}}\frac{D\,\dd D}{2\pi}
 w(D)J_n(DB)f_n(x,\bm k,D).
 \label{eq:second-hankel}
\end{align}
The window is part of the readout definition. Changing it changes spatial resolution and can produce oscillatory structure, so window dependence is reported as a resolution choice.

With $B$ in $\GeV^{-1}$, $W_w$ is dimensionless per unit $x$ and per natural-unit phase-space measure $\dd^2B\,\dd^2k$. When the plotted position axes are in fm, define
\begin{equation}
 W_w^{\rm fm}=\frac{W_w}{(\hbar c)^2},
 \qquad \hbar c=0.19733\GeV\,{\rm fm}.
 \label{eq:units}
\end{equation}
Then $W_w^{\rm fm}$ has units $\GeV^{-2}{\rm fm}^{-2}$ and is integrated with $\dd^2B_{\rm fm}\dd^2k$. Changing axis labels alone is insufficient if the color scale is described as a density.

An analytic transform check uses
\begin{equation}
 F(D,\phi)=A e^{-aD^2}(1+cD^2\cos 2\phi).
\end{equation}
The unrestricted transform is
\begin{equation}
 W(B,\phi_B)=\frac{A}{4\pi a}e^{-B^2/(4a)}
 \left[1-\frac{cB^2}{4a^2}\cos2\phi_B\right].
\end{equation}
This tests both the isotropic normalization and the negative sign of the quadrupole transform. Agreement with this analytic case validates the transform implementation; the transfer domain of a given model is validated separately, as in Sec.~\ref{sec:tomography}.

\section{Conditional supports and validation hierarchy}
\label{app:support}
For fixed state and evolution tables, $Q_{\rm ART}$ is convex in $p$. If $p_*$ is a candidate and $g=\nabla Q(p_*)$, convexity gives
\begin{equation}
 Q(p)\geq Q(p_*)+g^T(p-p_*).
\end{equation}
Minimizing this affine function over the product of simplices supplies a global lower bound for that finite quadratic problem. Target-support certificates additionally incorporate the compatibility constraint and are checked against attained primal endpoints. The nonlinear state step and alternative operator prescriptions are checked by their own tests.

Four levels of error are kept distinct: algebraic convention errors; representation and discretization errors; optimization errors; and model/source uncertainty. Tests proceed in that order. Exact symmetry and normalization tests precede projector and Fourier checks; refinement of the evolution grid precedes the kernel profiles; a new continuous integration domain requires interval checks, not just safe isolated nodes. The large-transfer pole discovery is the case in point: a new transform domain is validated by interval checks, and passing the old targets is necessary but not sufficient.

The calculations followed the staged protocol of Sec.~\ref{sec:architecture-validation}. The inputs and outputs of each stage, including the final fit, the evolution grids, the figure data, and the configurations of the phase-space evaluations, were recorded by content hash, so that each quoted number can be traced to the files that produced it. An AI coding agent (OpenAI Codex) was used to maintain this organization and these records and to run the validation checks on the code. The tomography calculations of Sec.~\ref{sec:tomography} are evaluations of the fixed calibrated model; no input covariance or fit result was changed in producing them.

\bibliography{references}
\end{document}